\documentclass[aps,prb,twocolumn,showpacs,preprintnumbers,amsmath,amssymb,superscriptaddress,10pt]{revtex4-1}
\usepackage{amsmath,amssymb}
\usepackage{bm}% bold math
\usepackage[utf8]{inputenc} % norske tegn
\usepackage{dcolumn}% Align table columns on decimal point
\usepackage{graphicx}% Include figure files
\usepackage{SIunits}
\usepackage{ae}
\usepackage{color} 
\usepackage{enumerate}

\usepackage{multirow}
\usepackage[caption=false]{subfig}
\usepackage{tikz}

\newcommand{\matrise}[1]{\begin{bmatrix} #1 \end{bmatrix}}
\newcommand{\diff}[0]{\text{d}}

\newcommand{\bigO}{O}
\newcommand{\re}{\mathrm{Re}\,}
\newcommand{\im}{\mathrm{Im}\,}

\newcommand{\vek}[1]{\boldsymbol{\mathbf{#1}}}
\newcommand{\vekh}[1]{\hat{{\boldsymbol{\mathbf{#1}}}}}
\renewcommand{\div}[0]{\nabla\cdot\vek}

\newcommand{\curl}{\nabla\times\vek}

\newcommand{\be}{\begin{equation}}
\newcommand{\ee}{\end{equation}}
\newcommand{\ba}{\begin{align}}
\newcommand{\ea}{\end{align}}
\newcommand{\e}[1]{\text{e}^{#1}}

\newcommand{\mc}[1]{\vek{\mathcal{#1}}}

\begin{document}
%\preprint{APS/123-QED}

\title{Four definitions of magnetic permeability for periodic metamaterials}% Force line breaks with\\

\author{Johannes Skaar}\email{johannes.skaar@its.uio.no}
\affiliation{Department of Technology Systems, University of Oslo, P.O. Box 70, NO-2027 Kjeller, Norway}
\author{Hans Olaf H\aa genvik}
\affiliation{Department of Electronic Systems, NTNU -- Norwegian University of Science and Technology, NO-7491 Trondheim, Norway}
\author{Christopher A. Dirdal}
\affiliation{SINTEF Digital, Microsystems and Nanotechnology, NO-0373 Oslo, Norway}

\date{\today}% It is always \today, today,

\begin{abstract}
We state and compare four different definitions of magnetic permeability for periodic, artificial media, or metamaterials. The connection between them, and properties in general, are discussed in detail, including causality, passivity, symmetry, asymptotic behavior, and origin dependence. The analysis is limited to metamaterials made from linear and nonmagnetic constituents.
\end{abstract}

%\pacs{78.20.Ci,42.70.-a,41.20.-q,42.25.Bs}% PACS, the Physics and Astronomy
\maketitle

\section{Introduction}\label{sec:intro}
In their famous textbook \cite{landau_lifshitz_edcm}, Landau and Lifshitz argue that the magnetic permeability ceases to have any physical meaning already at relatively low frequencies and above. The essence in their argument is that for high frequencies, the electric polarization current may become comparable or even larger than the current from the microscopic magnetization, contributing to the magnetic moment of the sample. The microscopic magnetization cannot therefore be interpreted as the total magnetic moment density.

For metamaterials, such as the split-ring resonator medium proposed by Pendry \cite{pendry1999}, the induced current in the inclusions is actually the main source of magnetism. By defining a macroscopic magnetization vector to describe a given part of the induced current, we obtain a definition of magnetic permeability which in principle can be used for all frequencies. However, this raises several questions. First of all, how should the induced current be decomposed into a magnetization term, electric polarization term and possibly other terms? Second of all, will the resulting permeability have the ``conventional'' properties that we expect for a permeability?

We limit the discussion to periodic media. Clearly, there is an infinite number of possibilities to decompose the induced current \cite{landau_lifshitz_edcm,vinogradov1999,vinogradov2002,silveirinha07,simovski09,alu11,Yaghjian2013374,dirdal18}; any transversal part of the induced current can be described both as a time-dependent, electric polarization term and a magnetization term. We will consider four possibilities: In the so-called Landau--Lifshitz formulation (Subsec. \ref{sec:LL}), all induced current is described by the electric polarization vector and therefore permittivity. Another natural and well known possibility is to define the magnetization as the magnetic moment density of the sample, using a fixed origin in each unit cell (Subsec. \ref{sec:multipole}). A variant of this approach was proposed by Yaghjian, Al\`u, and Silveirinha \cite{Yaghjian2013374}, using a decomposition of induced current due to Vinogradov and Aivazyan \cite{vinogradov1999} (Subsec. \ref{sec:VY}). A final possibility is to define the permeability to include ``as much as possible'' of the second order spatial dispersion of the Landau--Lifshitz permittivity. This approach was used by Landau, Lifshitz, and Pitaevskii \cite{landau_lifshitz_edcm}, and Silveirinha \cite{silveirinha07}, and is generalized here (Subsec. \ref{sec:TL}). How to construct other decompositions will be described briefly in Subsec. \ref{sec:other}.

Dependent on the particular decomposition, the resulting permeability gets more or less nonlocal (or dependent on wavenumber $\vek k$). However, at least for metamaterials which mimic natural magnetism, we expect that all four permeabilities coincide for low frequencies, and that they are local there. Nevertheless, to obtain a sufficiently large response, metamaterials are often used for relatively large frequencies where the lattice constant is comparable to the wavelength. In this region the permeabilities may differ (Sec. \ref{sec:numdisc}).

In Secs. \ref{sec:indcurr} and \ref{sec:numdisc} we will compare the different permeabilities, and discuss their properties, including causality/analyticity, passivity, symmetry, asymptotic behavior, and origin dependence. While some of these properties have been established previously, at least for certain permeabilities or with limited generality, the complete list, with associated proofs, is new to the best of our knowledge. In particular, we develop a rigorous framework where the source is treated as the proper input to the system, and obtain analyticity and invertibility for the tensor response function, and the Landau--Lifshitz permittivity tensor. This framework turns out to be useful to establish that all inverse permeabilities are causal (only one of them were known to be causal from Ref. \cite{Yaghjian2013374}). Furthermore, we determine the asymptotic behavior of the permeabilities. We also find analytically and numerically that all permeabilities may be different even for small $ka$, where $a$ is the lattice constant. This may appear surprising when comparing the definitions of magnetization in Subsecs. \ref{sec:multipole} and \ref{sec:VY}. Finally, a novel feature about the formulations is that even for nongyrotropic media, the magnetizations are allowed to depend on the longitudinal electric field. This is necessary to obtain a general treatment valid in the absence of symmetries.

Before reviewing the homogenization procedure, we will make a couple of definitions. The analysis happens in the frequency domain. The fields and parameters are clearly dependent on frequency in general; however for simplicity in notation we will usually not write this dependence explicitly. We use the standard notations $\bigO(k^n)$ and $\Theta(k^n)$ for the asymptotic behavior near zero or infinity; $\bigO(k^n)$ is used for expressions that is less than or equal to $C\cdot k^n$ ($C$ sufficiently large constant), while $\Theta(k^n)$ means expressions that tends to $C\cdot k^n$ for some constant $C$.

A time-domain function or distribution $f(t)$ is said to be \emph{causal} if it vanishes for $t<0$. A frequency-domain function $f(\omega)$ is said to be \emph{causal} if
\begin{enumerate}[(i)]
 \item $f(\omega)$ is analytic in an upper half-plane $\im\omega>\gamma$, where $\gamma$ is some real constant;
 \item $f(\omega)=\bigO(|\omega|^n)$ as $\omega\to\infty$ in this half-plane, for some integer $n$.
\end{enumerate}
This definition makes sense because of the following result from the theory of Laplace transforms \cite{doetsch74}:  
Any function $f(\omega)$ satisfying (i) and (ii) above can be represented as a Laplace transform of a causal time-domain function or distribution $f(t)$, setting the Laplace variable $s=-i\omega$.

\section{Homogenization}\label{sec:hom}
We consider a cubic periodic metamaterial. The metamaterial inclusions are assumed to be linear, nonmagnetic, passive and time-shift invariant. The microscopic, complex, relative permittivity in a unit cell will be denoted $\varepsilon(\vek r)$. The permittivity and permeability in vacuum are $\epsilon_0$ and $\mu_0$, respectively, and the vacuum light velocity is $c=1/\sqrt{\epsilon_0\mu_0}$. Angular frequency is denoted $\omega$. The microscopic Maxwell curl equations in the frequency domain are
\begin{subequations}\label{mmaxwell}
\begin{align}
\curl e(\vek r) &= i\omega\vek b(\vek r), \\
\frac{1}{\mu_0}\curl b(\vek r) &= -i\omega\epsilon_0\vek e(\vek r) + \vek j(\vek r)+\vek j_\text{ext}(\vek r),
\end{align}
\end{subequations}
with time dependence convention $\exp(-i\omega t)$. Here, $\vek j(\vek r)$ is the induced current density, which includes the ``bound'' current due to time-dependent, electric polarization density. Moreover, $\vek j_\text{ext}(\vek r)$ represents an external source current density, which can be expressed by an inverse Fourier transform
\be\label{spatFJ}
\vek j_\text{ext}(\vek r) = \frac{1}{(2\pi)^3}\int \vek J_\text{ext}(\vek k) \e{i\vek k\cdot\vek r}\diff^3 k.
\ee
To probe the metamaterial in the appropriate regime, it is natural to assume that the source is slowly varying over a unit cell size $a$, so that essentially, only $k$-values with $ka\ll 1$ contribute in the integral. However, this assumption is only necessary if we want our macroscopic fields to be true spatial averages (see the paragraph with Eqs. \eqref{macrok}-\eqref{averagefield} below). 

It is convenient to consider each spatial Fourier component in \eqref{spatFJ} separately, to enable the use of Floquet theory. Rather than \eqref{spatFJ}, we will therefore use a source \footnote{The dimension of $\vek J_\text{ext}(\vek k)$ in \eqref{spatJk} has an extra m$^{-3}$.}
\be\label{spatJk}
\vek j_\text{ext}(\vek r) = \vek J_\text{ext}(\vek k) \e{i\vek k\cdot\vek r}.
\ee
Then Floquet theory ensures that the fields can be written in the form
\begin{subequations}\label{floqeb}
\begin{align}
\vek e(\vek r)=\vek u_{\vek e}(\vek r,\vek k) \e{i\vek k\cdot\vek r}, \label{floqebe}\\
\vek b(\vek r)=\vek u_{\vek b}(\vek r,\vek k) \e{i\vek k\cdot\vek r}, \\
\vek j(\vek r)=\vek u_{\vek j}(\vek r,\vek k) \e{i\vek k\cdot\vek r},
\end{align}
\end{subequations}
where $\vek u_{\vek e}(\vek r,\vek k)$, $\vek u_{\vek b}(\vek r,\vek k)$, and $\vek u_{\vek j}(\vek r,\vek k)$ are periodic functions with periods equal to those of the material. Thus we can write
\be\label{per}
\vek u_{\vek e}(\vek r,\vek k)=\sum_{lmn}\vek E_{lmn}(\vek k)\e{i\vek b_{lmn}\cdot\vek r},
\ee
where $\vek b_{lmn}$'s are the reciprocal lattice vectors. In other words, the resulting field $\vek e(\vek r)$ contains a discrete Fourier spectrum, with a fundamental component
\be
\vek E(\vek k) \equiv \vek E_{000}(\vek k).
\ee
This component is the zeroth Fourier coefficient of the periodic function $\vek u_{\vek e}(\vek r,\vek k)$:
\begin{subequations}\label{floqebj}
\be\label{EmcE}
\vek E(\vek k) = \frac{1}{V}\int_V \vek u_{\vek e}(\vek r,\vek k)\diff^3r=\frac{1}{V}\int_V \vek e(\vek r)\e{-i\vek k\cdot\vek r}\diff^3r,
\ee
where $V$ denotes the volume of a unit cell. Note that \eqref{EmcE} is not a Fourier transform, as $\vek e(\vek r)$ is dependent on $\vek k$. Similarly, we have
\begin{align}
\vek B(\vek k) &= \frac{1}{V}\int_V \vek u_{\vek b}(\vek r,\vek k)\diff^3r=\frac{1}{V}\int_V \vek b(\vek r)\e{-i\vek k\cdot\vek r}\diff^3r, \\
\vek J(\vek k) &= \frac{1}{V}\int_V \vek u_{\vek j}(\vek r,\vek k)\diff^3r=\frac{1}{V}\int_V \vek j(\vek r)\e{-i\vek k\cdot\vek r}\diff^3r. \label{JmcJ}
\end{align}
\end{subequations}

As in Refs. \cite{silveirinha07,alu11,Yaghjian2013374}, we define the macroscopic field associated with the single-Fourier-component source as
\begin{subequations}\label{macrok}
\begin{align}
\mc E(\vek r)=\vek E(\vek k)\e{i\vek k\cdot\vek r}, \label{macrokE}\\
\mc B(\vek r)=\vek B(\vek k)\e{i\vek k\cdot\vek r}, \\
\mc J(\vek r)=\vek J(\vek k)\e{i\vek k\cdot\vek r}.
\end{align}
\end{subequations}
This definition, from the fundamental Floquet mode, can in principle be used for all $\vek k$ and $\omega$. Only when $ka\ll 1$, we can view the macroscopic fields as true spatial averages according to
\begin{subequations}\label{averagefield}
\begin{align}
   \mc E(\vek r) &= \int f(\mathbf{r}')\vek e(\mathbf{r-r}') \diff^3 r', \label{averagefielde}\\
   \mc B(\vek r) &= \int f(\mathbf{r}')\vek b(\mathbf{r-r}') \diff^3 r', \\
   \mc J(\vek r) &= \int f(\mathbf{r}')\vek j(\mathbf{r-r}') \diff^3 r'. \label{averagefieldj}
\end{align}
\end{subequations}
Here $f(\vek r)$ is a test function whose Fourier transform is negligible outside the first Brillouin zone, and normalized to unity for $\vek k=0$. The equivalence of \eqref{averagefield} and \eqref{macrok} under these conditions is established by Fourier transforming \eqref{averagefield} \cite{silveirinha09}.

Starting from the microscopic Maxwell equations \eqref{mmaxwell}, using \eqref{floqeb} and \eqref{per}, we can prove (see Appendix \ref{sec:maxwhom} for details):
\begin{subequations}\label{maxwellF}
\begin{align}
i\vek k\times\vek E(\vek k)-i\omega\vek B(\vek k) &=0, \label{faradayk}\\
\frac{1}{\mu_0} i\vek k\times\vek B(\vek k) + i\omega\epsilon_0\vek E(\vek k)-\vek J(\vek k) &= \vek J_\text{ext}(\vek k).
\end{align}
\end{subequations}
As will become clear in the next two paragraphs, Eqs. \eqref{maxwellF} should be viewed as the $\vek k$-space counterparts of Maxwell's equations for macroscopic fields $\mc E(\vek r)$, $\mc B(\vek r)$, and $\mc J(\vek r)$. They are \emph{not} the $\vek k$-space counterparts of the microscopic Maxwell equations.

In this work we will mostly use the single Fourier component source. However, we will now discuss the macroscopic field after superposition of the spatial Fourier components according to \eqref{spatFJ}. Rather than \eqref{macrokE} we then have the macroscopic field
\be\label{invF}
\mc E(\vek r)=\frac{1}{(2\pi)^3}\int\vek E(\vek k)\e{i\vek k\cdot\vek r}\diff^3k,
\ee
which is the inverse Fourier transform of the fundamental Floquet mode $\vek E(\vek k)$. The macroscopic fields $\mc B(\vek r)$ and $\mc J(\vek r)$ are expressed similarly. It is important to note that $\mc E(\vek r)\neq\vek e(\vek r)$ in general. Even for wavenumber spectra with $ka\ll 1$ the microscopic field $\vek e(\vek r)$ may vary rapidly in the unit cell, as described by the periodic function \eqref{per}. The operation \eqref{EmcE} picks only out the constant term in \eqref{per}, and the inverse Fourier transform \eqref{invF} is not able to restore the rapid variation.

By inverse Fourier transforming \eqref{maxwellF} we obtain the Maxwell equations for the macroscopic fields (or fundamental Floquet modes):
\begin{subequations}\label{maxwellfF}
\begin{align}
\nabla\times\mc E(\vek r)-i\omega\mc B(\vek r) &=0, \\
\frac{1}{\mu_0} \nabla\times\mc B(\vek r) + i\omega\epsilon_0\mc E(\vek r)-\mc J(\vek r) &= \vek j_\text{ext}(\vek r).
\end{align}
\end{subequations}
In principle, the Maxwell equations \eqref{maxwellF} and \eqref{maxwellfF} are valid for all $\omega$, and any spectra of $\vek k$'s. In other words, although it is natural to assume that $ka\ll 1$ for the contributing modes, such that the macroscopic fields are true spatial averages, we may in principle use the macroscopic fields for the entire $\vek k$ and $\omega$ spectrum, as long as we recall their meaning as fundamental Floquet modes. A natural question then is if the macroscopic fields have any physical significance for arbitrary $ka$. Indeed, it turns out that they can be used to calculate the work done by the source in each unit cell, provided the wavenumber spectrum is sufficiently narrow (Appendix \ref{sec:work}).

Note that in the presence of a source, $\omega$ and $\vek k$ are free parameters \cite{agranovich84,silveirinha07,alu11,Yaghjian2013374}, resulting from the Fourier decomposition of the source with respect to $t$ and $\vek r$. For example, the homogenized electric field is described in $(\omega, \vek k)$ space by the quantity $\vek E(\vek k)$, which is dependent on $\omega$ and $\vek k$ separately (the $\omega$-dependence is suppressed in the notation). For discussions on causality and asymptotic behavior we will hold $\vek k$ fixed and vary $\omega$. This corresponds to considering the frequency (or temporal) dependence of a single spatial Fourier component of the source, and the associated response. As seen below, this leads e.g. to a causal Landau--Lifshitz permittivity \cite{landau_lifshitz_edcm,agranovich84}. 

\section{Induced current}\label{sec:indcurr}
Now the big question is how to decompose the induced current density, to obtain a macroscopic permittivity, permeability, and possibly other parameters. In the most convenient and conventional case, we can express
\begin{subequations}\label{indcurrPM}
\begin{align}
\vek J(\vek k) &= -i\omega\vek P(\vek k) + i\vek k\times\vek M(\vek k), \\
\vek P(\vek k) &= \epsilon_0(\vek\epsilon-1)\vek E(\vek k), \\ 
\vek M(\vek k) &= \mu_0^{-1}(1-\vek\mu^{-1})\vek B(\vek k),
\end{align}
\end{subequations}
for some relative permittivity and permeability tensors $\vek\epsilon$ and $\vek\mu$ independent of $\vek k$. Then we have a local description of the constitutive relations. By defining auxiliary fields 
\begin{subequations}\label{vekDH}
\begin{align}
\vek D(\vek k) &= \epsilon_0\vek E(\vek k)+\vek P(\vek k), \\
\vek H(\vek k) &= \vek B(\vek k)/\mu_0-\vek M(\vek k),
\end{align}
\end{subequations}
Maxwell's equations \eqref{maxwellF} can be written
\begin{subequations}\label{maxwellFloc}
\begin{align}
i\vek k\times\vek E(\vek k)-i\omega\vek B(\vek k) &=0, \\
i\vek k\times\vek H(\vek k) + i\omega\vek D(\vek k) &= \vek J_\text{ext}(\vek k).
\end{align}
\end{subequations}
Transforming to the spatial domain,
\begin{subequations}\label{maxwellfFloc}
\begin{align}
\nabla\times\mc E(\vek r)-i\omega\mc B(\vek r) &=0, \\
\nabla\times\mc H(\vek r) + i\omega\mc D(\vek r) &= \vek j_\text{ext}(\vek r),
\end{align}
\end{subequations}
with
\begin{subequations}\label{mcDH}
\begin{align}
\mc D(\vek r) &= \epsilon_0\mc E(\vek r) + \mc P(\vek r) = \epsilon_0\vek\epsilon\mc E(\vek r), \label{mcDE}\\
\mc H(\vek r) &= \mc B(\vek r)/\mu_0-\mc M(\vek r) = \mu_0^{-1}\vek\mu^{-1}\mc B(\vek r),
\end{align}
\end{subequations}
and $\mc P(\vek r)$ and $\mc M(\vek r)$ are the inverse Fourier transform of $\vek P(\vek k)$ and $\vek M(\vek k)$, respectively. The equation set \eqref{maxwellfFloc} with \eqref{mcDH} is a local description of the electromagnetic fields.

In general, it is not always possible to express the induced current exactly as in \eqref{indcurrPM} with local constitutive parameters $\vek\epsilon$ and $\vek\mu$ (independent of $\vek k$). In Subsecs. \ref{sec:LL}-\ref{sec:TL} we will consider four possibilities how to decompose the induced current. All decompositions have appeared in previous literature, although the one in Subsec. \ref{sec:TL} has been generalized. In each subsection, we will discuss the properties of the different, resulting permeabilities. In Subsec. \ref{sec:other} we discuss how one can construct other decompositions and analyze their properties. 

We want Maxwell equations in the form \eqref{maxwellFloc} and \eqref{maxwellfFloc} to be valid in all cases; however with different expressions for the auxiliary fields $\vek D(\vek k)$ and $\vek H(\vek k)$. The strategy will be first to define a magnetization $\vek M(\vek k)$, then putting
\begin{subequations}\label{vekDHgen}
\begin{align}
\vek D(\vek k) &= \epsilon_0\vek E(\vek k) + \frac{\vek J(\vek k)- i\vek k\times\vek M(\vek k)}{-i\omega}, \label{Ddefgen}\\
\vek H(\vek k) &= \vek B(\vek k)/\mu_0-\vek M(\vek k).
\end{align}
\end{subequations}
Substituting \eqref{vekDHgen} into \eqref{maxwellFloc}, we recover \eqref{maxwellF}.

From now on, we will omit the $\vek k$ dependence in the notation, i.e., we will e.g. write $\vek J$ rather than $\vek J(\vek k)$. An exception is the Landau--Lifshitz permittivity in Subsec. \ref{sec:LL}, which always will be denoted $\vek\epsilon(\omega,\vek k)$, i.e., with arguments. Note that the fundamental fields, i.e., $\vek E$, $\vek B$, $\vek J$, and $\vek J_\text{ext}$, are the same in all formulations. We will often, without loss of generality, orient the coordinate system such that $\vek k$ points in the $\vekh x$-direction, i.e., $\vek k=k\vekh x$.

\subsection{Landau--Lifshitz (ll) formulation}\label{sec:LL}
In the Landau--Lifshitz formulation \cite{landau_lifshitz_edcm}, we describe all induced current in terms of a electric polarization density $\vek P^\text{ll}$:
\be
\vek J=-i\omega\vek P^\text{ll}.
\ee
This means that the magnetization is zero ($\vek M^\text{ll}=0$), and the permeability is trivial, $\vek\mu_\text{ll}=\vek I$. The displacement vector is $\vek D^\text{ll}=\epsilon_0\vek E+\vek P^\text{ll}$, or
\be\label{DLL}
\vek D^\text{ll}= \epsilon_0\vek E-\vek J/i\omega.
\ee
In a linear medium, there is a linear constitutive relation between $\vek D^\text{ll}$ and $\vek E$:
\be\label{DepsLL}
\vek D^\text{ll}=\epsilon_0\vek\epsilon(\omega,\vek k)\vek E.
\ee
This defines the Landau--Lifshitz permittivity $\vek\epsilon(\omega,\vek k)$. We note that the constitutive relations are described in the form of a single parameter, $\vek\epsilon(\omega,\vek k)$. Considering terms up to second order in $k$,
\be\label{LLepsgamma2}
\epsilon_{ij}(\omega,\vek k) - \delta_{ij} = \chi_{ij} + \alpha_{ikj}k_k/\epsilon_0 + \beta_{iklj}k_kk_l c^2/\omega^2,
\ee
for some tensors $\chi_{ij}$, $\alpha_{ikj}$, and $\beta_{iklj}$, independent of $\vek k$. In \eqref{LLepsgamma2} summation over repeated indices is implied. In the presence of strong spatial dispersion, where higher order terms are not negligible, we let the $\beta_{iklj}k_kk_l c^2/\omega^2$ term absorb the remainder. For such media the $\beta_{iklj}$ tensor gets dependent on $\vek k$.

Maxwell's equations \eqref{maxwellF} take the form
\begin{subequations}\label{maxwellLL00}
\begin{align}
i\vek k\times\vek E-i\omega\vek B &=0, \\
\frac{1}{\mu_0} i\vek k\times\vek B + i\omega\epsilon_0\vek\epsilon(\omega,\vek k)\vek E &= \vek J_\text{ext}.
\end{align}
\end{subequations}
By eliminating $\vek B$, we obtain
\be\label{maxwmatr00}
\left(k^2\vek I_\perp-\frac{\omega^2}{c^2}\vek\epsilon(\omega,\vek k)\right)\vek E = i\omega\mu_0\vek J_\text{ext},
\ee
with $\vek I_\perp=\vek I - \vek k\vek k/k^2$, where $\vek I$ is the identity, or
\be\label{Iperpx}
\vek I_\perp = \matrise{0 & 0 & 0 \\ 0 & 1 & 0 \\ 0 & 0 & 1},
\ee
expressed in a coordinate system where $\vek k=k\vekh x$. The matrix in the brackets in \eqref{maxwmatr00} can be inverted (Appendix \ref{sec:anal}), to obtain an input-output relation
\be\label{ERJresp00}
\vek E = \vek G(\omega,\vek k)\vek J_\text{ext},
\ee
where $\vek G(\omega,\vek k)$ is a (matrix) response function given by
\be\label{Rinv00}
\vek G(\omega,\vek k)^ {-1} = (i\omega\mu_0)^{-1}\left(k^2\vek I_\perp-\frac{\omega^2}{c^2}\vek\epsilon(\omega,\vek k)\right).
\ee

For an isotropic medium, the permittivity tensor can be written
\be
\vek\epsilon(\omega,\vek k) = \matrise{\epsilon_\parallel & 0 & 0 \\ 0 & \epsilon_\perp & 0 \\ 0 & 0 & \epsilon_\perp},
\ee
for a longitudinal $\epsilon_\parallel$ and transversal $\epsilon_\perp$ permittivity, respectively. In this case the response function $\vek G(\omega,\vek k)$ becomes $G(\omega,k)=1/i\omega\epsilon_0\epsilon_\parallel$ or 
\be\label{resp100}
G(\omega,k) = \frac{i\omega\mu_0}{k^2-\frac{\omega^2}{c^2}\epsilon_\perp}, 
\ee
dependent on the direction of the source $\vek J_\text{ext}$.

For each $\vek k$, we have, due to passivity and causality (Appendix \ref{sec:anal}):
\begin{subequations}\label{epsGprop}
 \begin{align}
  & \vek G(\omega,\vek k) \text{ analytic for }\im\omega>0 \text{ and fixed $\vek k$} , \\
- & \vek G(\omega,\vek k)^{-1} - \vek G(\omega,\vek k)^{-1\dagger} \text{ positive definite}, \\
  & \det\vek G(\omega,\vek k) \neq 0 \text{ for } \im\omega>0, \label{Rinvertible00}\\
  & \det\vek G(\omega,\vek k)^{-1} \neq 0 \text{ for } \im\omega>0, \\
  & \vek\epsilon(\omega,\vek k) \text{ analytic for }\im\omega>0 \text{ and fixed $\vek k$}, \\
   - & i\omega[\vek\epsilon(\omega,\vek k) - \vek\epsilon(\omega,\vek k)^\dagger] \text{ positive semidefinite}. \label{epspass0}
 \end{align}
\end{subequations}
 Here $\dagger$ denotes Hermitian conjugate (transpose and complex conjugate). For \eqref{epspass0} we have assumed real $\omega$ and $\vek k$, as is the case for Fourier decomposition of the fields (Sec. \ref{sec:hom}). If the Fourier integrals in $\omega$ and $\vek k$ are deformed into the complex plane, the permittivity satisfies \eqref{passivitygen} rather than \eqref{epspass0}.

For reciprocal metamaterial inclusions, we have
\begin{subequations}\label{recGeps}
 \begin{align}
  & \vek G^\text{T}(\omega,-\vek k) = \vek G(\omega,\vek k), \\
  & \vek\epsilon^\text{T}(\omega,-\vek k) = \vek\epsilon(\omega,\vek k), \label{receps}
 \end{align}
\end{subequations}
where the superscript ``T'' denotes transpose. From \eqref{Rinv00} the two equations in \eqref{recGeps} are equivalent. The symmetry relation \eqref{receps} is well known in literature \cite{landau_lifshitz_edcm,agranovich84}; a proof can be found in \cite{Yaghjian2013374}.
 
For nongyrotropic media, we have $\vek\epsilon(\omega,-\mathbf{k})=\vek\epsilon(\omega,\mathbf{k})$ \cite{landau_lifshitz_edcm,agranovich84}. This will be the case if there is a center of symmetry in the medium. Then the odd-order term in \eqref{LLepsgamma2} vanishes,
\be\label{alpha0} 
\alpha_{ikj}=0.
\ee

The asymptotic behavior of $\vek\epsilon(\omega,\vek k)$ as $\omega\to\infty$ can be viewed in two different ways. In principle, for sufficiently large frequencies the permittivities of the inclusions and host medium tend to unity \cite{landau_lifshitz_edcm}; thus eventually $\vek\epsilon(\omega,\vek k)\to\vek I$. Nevertheless, in some cases it can be convenient to describe the asymptotic behavior as $\epsilon(\omega,\vek k)\to\text{const}$, where the constant tensor limit can be different from identity. This may be the case e.g if the permittivities of the inclusions and the host medium are considered nondispersive in the frequency range of interest. 

With either of these asymptotic behaviors, the tensors $\vek\epsilon(\omega,\vek k)$, $\vek G(\omega,\vek k)^{-1}$, and $\vek G(\omega,\vek k)$ are causal functions. This follows from the definition of a causal function in Sec. \ref{sec:intro}, and \eqref{Rinv00} and \eqref{epsGprop}.

\subsection{Multipole decomposition}\label{sec:multipole}
The traditional way to decompose the induced current, is by multipole expansion \cite{bladel07,alu11,dirdal18}. Consider the unit cell that contains the origin. Using 
\be\label{expandexp}
\exp(-i\vek k\cdot\vek r)=1-i\vek k\cdot\vek r-(\vek k\cdot\vek r)^2/2+\Theta((\vek k\cdot\vek r)^3),
\ee
we obtain from \eqref{JmcJ} to second order in $k$:
\begin{align}
& \vek J
= \frac{1}{V}\int_V \mathbf{j} \e{-i\vek k\cdot\vek r}\diff^3r \label{eq:AvPMul} \\
&= \frac{1}{V} 
\cdot\left(\int_V \mathbf{j}\diff^3r - i\vek k\cdot\int_V \vek r\mathbf{j}\diff^3r - \frac{1}{2}\int_V (\vek k\cdot\vek r)^2 \mathbf{j} \diff^3r \right) \nonumber\\
&\equiv -i\omega\vek P +i\vek k\times\vek M - \omega\vek k\cdot\vek Q/2 - i\omega\vek R, \label{eq:Multipole}
\end{align}
where
\begin{subequations}\label{eq:MultipoleVectors}
\begin{align}
\vek P &= \frac{1}{-i\omega V}\int_V \mathbf{j}\diff^3 r, \\
\vek M &= \frac{1}{2V}\int_V \vek r\times\vek j\diff^3 r, \label{eq:MultipoleVectorsMagn} \\
\vek Q &= \frac{1}{-i\omega V}\int_V (\vek r\mathbf{j}+\mathbf{j}\vek r)\diff^3 r, \\
\vek R &= \frac{1}{2i\omega V}\int_V (\vek k\cdot\vek r)^2\mathbf{j}\diff^3 r. \label{eq:RTerm}
\end{align}
\end{subequations}
Here we have decomposed the tensor $\vek r\mathbf{j}$ into its antisymmetric and symmetric parts,
\begin{align}\label{eq:DecomSymAndAntiSym}
\vek k\cdot\vek r\mathbf{j} &= \vek k\cdot(\vek r\mathbf{j}-\mathbf{j}\vek r)/2+\vek k\cdot(\vek r\mathbf{j}+\mathbf{j}\vek r)/2 \nonumber\\
&= -\vek k\times\vek r\times\mathbf{j}/2 +\vek k\cdot(\vek r\mathbf{j}+\mathbf{j}\vek r)/2.
\end{align}
In addition to the polarization vector $\vek P$, magnetization vector $\vek M$, and quadrupole tensor $\vek Q$, the extra term $\vek R$ includes electric octupole and magnetic quadrupole. All these multipole terms are dependent on $\vek k$ although not explicitly specified.

A convenient feature of the multipole decomposition is that the terms have a clear physical interpretation. In particular, $\vek M$ quantifies the amount of circulating, induced currents. For example, if a 2d metamaterial unit cell consists of a cylinder inclusion with a circular symmetric current in the azimuthal direction, we obtain $\vek P=0$, $\vek Q=0$, and $\vek R=0$, while $\vek M$ is nonzero.

From $\vek M$ we define, as usual,
\be\label{HMmult}
\vek H=\vek B/\mu_0-\vek M.
\ee
The remaining terms in \eqref{eq:Multipole} go into the displacement vector, according to \eqref{Ddefgen} \footnote{Alternatively, the electric octupole--magnetic quadrupole term $\vek R$ could be split such that the magnetic quadrupole is included into \eqref{HMmult}.}:
\be\label{Dmp}
\vek D = \epsilon_0\vek E+\vek P-i\vek k\cdot\vek Q/2+\vek R.
\ee

In a linear medium, we can write the associated constitutive relations
\begin{subequations}\label{eq:2ndOrderkExp}
\begin{align}
P_i &= \epsilon_0\chi_{ij} E_j + \xi_{ikj}k_kE_j + \eta_{iklj}k_kk_lE_j/(\mu_0\omega^2), \label{P2o}\\
M_{i} &= \omega\zeta_{ij}E_j + \nu_{ilj}k_lE_j/(\mu_0\omega), \label{eq:MExp} \\
Q_{ik} &= 2i\sigma_{ikj}E_j + 2i\gamma_{iklj}k_lE_j/(\mu_0\omega^2), \label{eq:QuadExp} \\
R_i &= \psi_{iklj}k_k k_l E_j/(\mu_0\omega^2),
\end{align}
\end{subequations} 
for some tensors $\chi_{ij}$, $\xi_{ikj}$, $\eta_{iklj}$, $\sigma_{ikj}$, $\gamma_{iklj}$, $\psi_{iklj}$, and pseudotensors $\zeta_{ij}$ and $\nu_{ilj}$. Treating the (pseudo-)ten\-sors as Taylor coefficients independent of $\vek k$, we have included the necessary orders of $k$ such that $\vek J$ is second order when substituting in \eqref{eq:Multipole}. We can consider higher order spatial dispersion by letting the highest order term in \eqref{eq:2ndOrderkExp} take care of the remainder. For example, in \eqref{eq:MExp} this will lead to a $\nu_{ilj}$ which is dependent on $\vek k$.

From Faraday's law $\vek B=\vek k\times\vek E/\omega$, we note that any dependence on $\vek B$ is taken care of by the $\vek k$ dependent terms in \eqref{eq:2ndOrderkExp}. For later convenience we have included certain $\vek k$ independent quantities (such as $\mu_0\omega^2$) in the tensor elements. Magneto-electric coupling is taken into account in terms of $\xi_{ikj}$ and $\zeta_{ij}$.  

We are interested in the magnetization \eqref{eq:MExp}. Choosing coordinate system such that $\vek k=k\vekh x$, we can write
\begin{align}
M_{i} &= \omega\zeta_{ij}E_j + k\nu_{i1j}E_j/(\mu_0\omega) \label{eq:MExp2}\\
 &= \omega\zeta_{ij}E_j + k\nu_{i11}E_1/(\mu_0\omega) + \mu_0^{-1} (1-\vek\mu^{-1})_{ij}B_j, \nonumber 
\end{align}
with
\be\label{munu}
1-\vek\mu^{-1} = \matrise{-\nu_{213} & \nu_{212} \\ -\nu_{313} & \nu_{312} }.
\ee
Here $\vek\mu^{-1}$ is identified as an inverse permeability, resulting from the magnetization $\vek M$ defined as the averaged magnetic moment density \eqref{eq:MultipoleVectorsMagn}. Note that in the coordinate system where $\vek k=k\vekh x$, the inverse permeability is described as $2\times 2$. The reason for this is that $\vek B$ is transversal (i.e., $B_1=0$), and that only the transversal part of $\vek M$ contributes to $\vek J$ by \eqref{eq:Multipole}. In an arbitrary coordinate system, \eqref{munu} can be written
\be\label{munuarbcoord}
(1-\vek\mu^{-1})_{im} = \varepsilon_{mkj}\frac{k_kk_l}{k^2}\nu_{ilj},
\ee
where $\varepsilon_{mkj}$ is the Levi-Civita symbol. This means that $1-\vek\mu^{-1}$ is a tensor.

We will now compare the Landau--Lifshitz formulation and the multipole decomposition. By eliminating $\vek D$ from \eqref{DLL} and \eqref{DepsLL}, and comparing with \eqref{eq:Multipole}, we obtain
\be
\epsilon_0\vek\epsilon(\omega,\vek k)\vek E = \epsilon_0\vek E + \vek P -\vek k\times\vek M/\omega -i \vek k\cdot\vek Q/2 + \vek R.
\ee
Using the constitutive relations \eqref{eq:2ndOrderkExp} this gives
\begin{align}
&\epsilon_{ij}(\omega,\vek k) - \delta_{ij} = \chi_{ij} + \left(\xi_{ikj} + \sigma_{ikj} - \varepsilon_{ikm}\zeta_{mj}\right)k_k/\epsilon_0   \nonumber \\
&+ \left(\gamma_{iklj}+\psi_{iklj}+\eta_{iklj}-\varepsilon_{ikm}\nu_{mlj}\right) k_kk_l c^2/\omega^2. \label{epsLLparm}
\end{align}  
Comparing \eqref{LLepsgamma2} and \eqref{epsLLparm}, and noting that $\beta_{iklj}$, $\psi_{iklj}$, and $\eta_{iklj}$ can be taken to be symmetric in $k$ and $l$, we have
\begin{subequations}
\begin{align}
& \alpha_{ikj} = \xi_{ikj} + \sigma_{ikj} - \varepsilon_{ikm}\zeta_{mj}, \\
& \beta_{iklj} = \psi_{iklj} + \eta_{iklj} + \frac{\gamma_{iklj}+\gamma_{ilkj}}{2} - \frac{\varepsilon_{ikm}\nu_{mlj}+\varepsilon_{ilm}\nu_{mkj}}{2}.
\end{align}  
\end{subequations}

For nongyrotropic media, if there is a center of symmetry in the medium, we can take the center of the unit cell to be the center of symmetry. For $\vek k=0$, from symmetry and \eqref{eq:MultipoleVectors}, it follows that $\vek M=0$ and $\vek Q=0$. This means that $\zeta_{ij}=0$, $\sigma_{ikj}=0$, and from \eqref{alpha0}, $\xi_{ikj}=0$.

In other words, for nongyrotropic media, $\vek M$ and $\vek Q$ contain only first order terms in $k$, which means that all terms in \eqref{eq:Multipole} except $\vek P$ are second order in $k$. This means that the electric octupole--magnetic quadrupole term $\vek R$ can be of the same order of magnitude as the magnetization and quadrupole terms \cite{dirdal18}. Thus, when concerned with the magnetic response, the $\vek R$-term and $\vek Q$ should in general be taken into account in addition to $\vek M$.

Even when considering an asymptotic behavior of the microscopic permittivity $\varepsilon(\vek r)\to 1$ as $\omega\to\infty$, it turns out that for fixed $\vek k$, we have $\vek\mu^{-1}\not\to\vek I$ in general \cite{dirdal18d}. An asymptotic value different from identity does not violate causality, as $\vek\mu^{-1}\to\vek I$ is only required for eigenmode propagation where $\omega$ and $\vek k$ are connected. Even though the asymptotic behavior for fixed $\vek k$ may have limited direct physical importance, it has implications for the Kramers--Kronig relations, being formulated for fixed $\vek k$. The asymptotic behavior of $\vek\mu$ is found as follows. The asymptotic behavior of any microscopic permittivity is of the form \cite{landau_lifshitz_edcm}
\be\label{varepsas}
\varepsilon(\vek r) = 1 - \frac{\omega_\text{p}^2(\vek r)}{\omega^2} + \bigO(\omega^{-3}),
\ee
where $\omega_\text{p}(\vek r)$ is the plasma frequency. As $\omega\to\infty$ the fields will tend to those we would have if the metamaterial were replaced by vacuum. Thus we can write
\begin{subequations}
\label{microje}
\begin{align}
\vek e(\vek r) &= \vek E\exp(i\vek k\cdot\vek r)+\vek f(\vek r), \\
\vek j(\vek r) &= -i\omega\epsilon_0[\varepsilon(\vek r)-1][\vek E\exp(i\vek k\cdot\vek r)+\vek f(\vek r)], \label{microj} 
\end{align}
\end{subequations}
for some $\vek f(\vek r)$, with 
\be\label{fasympt}
\vek f(\vek r)\to 0 \text{ as } \omega\to\infty.
\ee
Here we have assumed a source such that $\vek E$ is independent of $\omega$ for large frequencies (this condition can be removed). Having an expression for $\vek j(\vek r)$, it is straightforward to obtain $\vek M$ by \eqref{eq:MultipoleVectorsMagn}:
\begin{align}\label{MEomeginfmp}
\vek M &= \frac{i\omega\epsilon_0}{2V}\vek E\times\int_V \vek r \left[\varepsilon(\vek r)-1\right]\e{i\vek k\cdot\vek r}\diff^3r \nonumber\\
 &- \frac{i\omega\epsilon_0}{2V} \int \vek r\times\vek f(\vek r)[\varepsilon(\vek r)-1]\diff^3r.
\end{align}
According to \eqref{varepsas} and \eqref{fasympt}, the last term in \eqref{MEomeginfmp} tends to zero faster than $\omega^{-1}$. Comparing to \eqref{eq:MExp}, this means that the term will not contribute to $\nu_{ilj}$ in the limit $\omega\to\infty$. The first term in \eqref{MEomeginfmp} can be written
\be\label{1term}
\frac{-i\epsilon_0}{2\omega V}\vek E\times\int_V \vek r\,\omega_\text{p}^2(\vek r)\e{i\vek k\cdot\vek r}\diff^3r + \bigO(\omega^{-2}).
\ee
The integral in \eqref{1term} is clearly nonzero in general. Then \eqref{1term} is $\Theta(\omega^{-1})$, which by \eqref{eq:MExp} means that $\vek\nu\not\to 0$. We therefore find that
\be\label{asymptmumult}
1-\vek\mu^{-1} = \bigO(1) \text{ as }\omega\to\infty, \text{ for fixed }\vek k,
\ee
and in general, $\vek\mu^{-1}\not\to\vek I$. 

For diagonal $\vek\mu$ it is straightforward to find examples where $\im\vek\mu$ is both positive and negative, depending on the frequency (see Sec. \ref{sec:numdisc}). This is not a violation of passivity; it is just an indication of the phase relationship between the magnetization and the macroscopic field in the unit cell. The fundamental passivity condition is only that the Landau--Lifshitz permittivity satisfies \eqref{epspass0}.

We will now consider the causality and analyticity of the inverse permeability. Note that $\vek E$ is the same in all formulations, so we can use the Landau--Lifshitz formulation to express
\be
\vek E = \vek G(\omega,\vek k)\vek J_\text{ext}, \label{EJLL}\\  
\ee
with a response function $\vek G(\omega,\vek k)$, as in \eqref{ERJresp00}. According to \eqref{Rinvertible00}, $\vek G(\omega,\vek k)$ is invertible in the upper half-plane $\im\omega>0$. Hence, we can choose $\vek J_\text{ext}$ such that only a single component of $\vek E$ is nonzero, say, $E_j$, and such that $E_j$ is any analytic and causal function. The required $\vek J_\text{ext}$ is analytic in the upper half-plane, from the analyticity of $\vek G(\omega,\vek k)^{-1}$. Taking the asymptotic behavior of $\vek G^{-1}(\omega,\vek k)$ as $\omega\to\infty$ into account, the required $\vek J_\text{ext}$ is realizable as a causal source. 

We have from \eqref{eq:MExp2} that
\be
\label{eq:MExpn}
M_i = \omega\zeta_{ij}E_j + k\nu_{i1j}E_j/(\mu_0\omega),
\ee
where now, only a single component $E_j$ is nonzero. Clearly the microscopic, induced current $\vek j$ is causal, since it is causally related to the source. Thus $M_i$, as given by \eqref{eq:MultipoleVectorsMagn}, is causal. Putting $k=0$ in \eqref{eq:MExpn}, and remembering that $E_j$ is any causal function, it follows that $\zeta_{ij}$ is analytic in the upper half-plane. By letting $k\neq 0$, we find that $\nu_{i1j}$ is analytic there, since $M_i$ and $\omega\zeta_{ij}E_j$ are. From \eqref{munu} we conclude that $\vek\mu^{-1}$ is analytic in the upper half-plane. Moreover, taking \eqref{asymptmumult} into account, $\vek\mu^{-1}$ is causal. Writing $\vek\mu^{-1}(\omega,\vek k)\to\vek\mu^{-1}(\infty,\vek k)$, we can establish Kramers--Kronig relations \eqref{KK} for $\vek\chi(\omega,\vek k)\equiv \vek\mu^{-1}(\omega,\vek k)-\vek\mu^{-1}(\infty,\vek k)$ [23].

It is also possible to combine $\zeta_{ij}$ and $\mu^{-1}_{ij}$ into a single, inverse permeability tensor \cite{Yaghjian2013374}, and consider its causality. In a coordinate system where $\vek k=k\vekh x$, Faraday's law \eqref{faradayk} becomes $E_2=B_3\omega/k$ and $E_3=-B_2\omega/k$. We can then express \eqref{eq:MExp2} as
\begin{align}
M_{i} &= \omega\zeta_{i1}E_1 + k\nu_{i11}E_1/(\mu_0\omega) \\
 &+ \omega^2\zeta_{i2}B_3/k - \omega^2\zeta_{i3}B_2/k + \mu_0^{-1} (1-\vek\mu^{-1})_{ij}B_j, \nonumber
\end{align}
or
\be\label{eq:MExp3}
M_{i} = \omega\zeta_{i1}E_1 + k\nu_{i11}E_1/(\mu_0\omega) + \mu_0^{-1} (1-\tilde{\vek\mu}^{-1})_{ij}B_j 
\ee
with the modified inverse permeability
\be\label{mupcomb}
\tilde{\vek\mu}^{-1} = \vek\mu^{-1} - \frac{\mu_0\omega^2}{k}\matrise{-\zeta_{23} & \zeta_{22} \\ -\zeta_{33}  & \zeta_{32}}.
\ee
In the two previous paragraphs we found that $\vek\mu^{-1}$ and $\zeta_{ij}$ are analytic in the upper half-plane; thus so is $\tilde{\vek\mu}^{-1}$.

It is interesting to note that all (pseudo-)tensor elements in \eqref{eq:2ndOrderkExp} are analytic in the upper half-plane. This is seen as follows. First, recall from \eqref{EJLL} and \eqref{Rinvertible00} that the source can be chosen such that only a single component of the electric field, say $E_j$, is nonzero, and such that $E_j$ is any analytic function. Also, $P_i$, $M_i$, $Q_{ik}$, and $R_i$ are analytic, since they are given by the induced, microscopic current through \eqref{eq:MultipoleVectors}. We now apply the general result in Appendix \ref{sec:analtensor} to the expansions \eqref{eq:2ndOrderkExp}, with the result that all (pseudo-)tensor elements in \eqref{eq:2ndOrderkExp} are analytic in the upper half-plane.

Finally we note the well known fact \cite{raab_lange} that in general, the multipole quantities are dependent on the choice of origin. We have assumed that the origin is inside the unit cell $V$, but we are free to move the origin inside the cell. Substituting $\vek r=\vek r'+\vek r_0$ in \eqref{eq:AvPMul}, and expanding the exponential $\exp(-i\vek k\cdot\vek r')$ give
\be\label{multJp}
\vek J = \e{-i\vek k\cdot\vek r_0}\left( -i\omega\vek P +i\vek k\times\vek M' - \omega\vek k\cdot\vek Q'/2 - i\omega\vek R' \right),
\ee
with
\begin{subequations}\label{eq:MultipoleVectorsp}
\begin{align}
\vek M' &= \frac{1}{2V}\int_V \vek r'\times\vek j\diff^3 r, \label{eq:MultipoleVectorsMagnp} \\
\vek Q' &= \frac{1}{-i\omega V}\int_V (\vek r'\mathbf{j}+\mathbf{j}\vek r')\diff^3 r, \\
\vek R' &= \frac{1}{2i\omega V}\int_V (\vek k\cdot\vek r')^2\mathbf{j}\diff^3 r.
\end{align}
\end{subequations}
By changing $\vek r_0$, the different multipole quantities will change; however such that the sum of contributions to the induced current (right-hand side of \eqref{multJp}) is constant. Since
\be
\vek M' = \vek M + \frac{i\omega\vek r_0}{2}\times\vek P,
\ee
we have $\vek M'\approx\vek M$ when $\omega a P \ll M$.

Since the magnetization vector is dependent on the choice of origin, so is the resulting $\vek\mu$ in general. This dependence is not only a consequence of the difference between $\vek M'$ and $\vek M$, but also the exponential factor $\exp(-i\vek k\cdot\vek r_0)\approx 1-i\vek k\cdot\vek r_0$ in \eqref{multJp}. This factor will mix the $\Theta(1)$ and $\Theta(k)$ terms in \eqref{eq:MExpn} in the presence of magnetoelectric coupling ($\zeta_{ij}\neq 0$).

\subsection{Vinogradov--Yaghjian (vy) decomposition}\label{sec:VY}
In Vinogradov and Aivazyan \cite{vinogradov1999} the microscopic current is decomposed into three terms:
\be\label{VAdecomp}
\vek j = -\vek r\,\div j + \frac{1}{2}\nabla\times(\vek r\times\vek j) + \frac{1}{2}\nabla\cdot (\vek r\vek j + \vek j\vek r).
\ee
Eq. \eqref{VAdecomp} can be verified by straightforward calculation. The microscopic current satisfies continuity $\div j=i\omega\varrho$, where $\varrho$ is the microscopic induced charge density. Yaghjian, Al\`u, and Silveirinha \cite{Yaghjian2013374} suggested to decompose the macroscopic induced current by substituting \eqref{VAdecomp} into \eqref{JmcJ}, resulting in
\be\label{Vinexp}
\vek J = -i\omega\vek P^{\text{vy}} + i\vek k\times\vek M^{\text{vy}} + \omega\vek k\cdot\vek Q^{\text{vy}}/2,
\ee
where
\begin{subequations}\label{PMQ}
\begin{align}
\vek P^{\text{vy}} &= \frac{1}{V}\int_V\varrho(\vek r)\vek r\e{-i\vek k\cdot\vek r}\diff^3r, \label{PMQP}\\
\vek M^{\text{vy}} &= \frac{1}{2V}\int_V\vek r\times\vek j(\vek r)\e{-i\vek k\cdot\vek r}\diff^3r, \label{PMQM}\\
\vek Q^{\text{vy}} &= -\frac{1}{i\omega V}\int_V(\vek j\vek r + \vek r\vek j)\e{-i\vek k\cdot\vek r}\diff^3r.
\end{align}
\end{subequations}
The integrals are over the unit cell containing the origin. To obtain \eqref{Vinexp} it is assumed that the boundaries of the unit cells lie in free space. Eq. \eqref{Vinexp} is not a multipole expansion, due to the factor $\exp(-i\vek k\cdot\vek r)$ in the integrands of \eqref{PMQ}. All induced current is described by the three terms in \eqref{Vinexp}, as opposed to a multipole expansion with an infinite number of terms. Note that the sign of the ``quadrupole'' term $\omega\vek k\cdot\vek Q^{\text{vy}}/2$ is opposite of that resulting from a conventional multipole expansion \eqref{eq:Multipole}.

From the magnetization $\vek M^{\text{vy}}$, we can define a permeability exactly as in Subsec. \ref{sec:multipole}. From a constitutive relation
\be\label{constitutiveMVY}
M_i^{\text{vy}} = \omega\zeta^{\text{vy}}_{ij}E_j + \nu^{\text{vy}}_{ilj}k_lE_j/(\mu_0\omega),
\ee
set
\be\label{munuarbcoordVY}
\left(1-\vek\mu_{\text{vy}}^{-1}\right)_{im} = \varepsilon_{mkj}\frac{k_kk_l}{k^2}\nu_{ilj}^{\text{vy}},
\ee
or
\be\label{munuVY}
1-\vek\mu_{\text{vy}}^{-1} = \matrise{-\nu_{213}^{\text{vy}} & \nu_{212}^{\text{vy}} \\ -\nu_{313}^{\text{vy}} & \nu_{312}^{\text{vy}} }
\ee
in a coordinate system where $\vek k=k\vekh x$. (Alternatively, as in Ref. \cite{Yaghjian2013374} and in \eqref{mupcomb}, we can define a new permeability $\tilde{\vek\mu}_{\text{vy}}$ by combining $\vek\mu_\text{vy}$ and $\vek\zeta^\text{vy}$ into a single tensor.)

The asymptotic behavior of $\vek\mu_\text{vy}^{-1}$ turns out to be different from that of $\vek\mu^{-1}$ in Subsec. \ref{sec:multipole}. Substituting \eqref{microj} into \eqref{PMQM}:
\begin{align}\label{MEomeginf}
\vek M^\text{vy} &= \frac{i\omega\epsilon_0}{2V}\vek E\times \int_V \vek r\left[\varepsilon(\vek r)-1\right]\diff^3r \nonumber\\
 &- \frac{i\omega\epsilon_0}{2V} \int\vek r\times\vek f(\vek r)[\varepsilon(\vek r)-1]\e{-i\vek k\cdot\vek r}\diff^3r.
\end{align}
The first integral is independent of $\vek k$, and cannot therefore contribute to the last term in \eqref{constitutiveMVY}. The second term in \eqref{MEomeginf} tends to zero faster than $\omega^{-1}$ (see \eqref{varepsas} and \eqref{fasympt}), and leads to a $\nu^\text{vy}_{ilj}$ that tends to zero. We therefore find that 
\be\label{asymptmuvy}
\vek\mu_\text{vy}^{-1}\to\vek I \text{ as }\omega\to\infty.
\ee

The definition of $\vek M^\text{vy}$ in \eqref{PMQM} can be used to prove that $\vek\mu_\text{vy}^{-1}$ is analytic in the upper half-plane $\im\omega>0$, using the exact same method as in Subsec. \ref{sec:multipole}. This result is already known from \cite{Yaghjian2013374}. Taking \eqref{asymptmuvy} into account, we conclude that $\vek\mu_\text{vy}^{-1}$ is causal for each, fixed $\vek k$.

The connection between the constitutive parameters for $\vek P^{\text{vy}}$, $\vek M^{\text{vy}}$, $\vek Q^{\text{vy}}$, and the Landau--Lifshitz permittivity can be obtained directly from \eqref{epsLLparm} by setting $\psi_{iklj}=0$ (and adding superscripts ``vy'').

At first sight, the multipole quantities in \eqref{eq:MultipoleVectors} and in \eqref{PMQ} seem to be quite similar; the difference is only a factor $\exp(-i\vek k\cdot\vek r)$ in the integrands. The connection between the multipole quantities can be established by expanding the exponential \eqref{expandexp}. Note that since we are interested in magnetic effects, which are known to be a second order $\Theta(k^2)$ effect in the Landau-Lifshitz permittivity, we include terms for the induced current up to order $\Theta(k^2)$. Expressing $i\omega\varrho=\nabla\cdot{\vek j}$ and using integration by parts, we obtain from \eqref{PMQP}:
\be
-i\omega\vek P^{\text{vy}} = \frac{1}{V}\int_V\vek j\e{-i\vek k\cdot\vek r}\diff^3r - \frac{i\vek k\cdot}{V}\int_V\vek j\vek r\e{-i\vek k\cdot\vek r}\diff^3r.
\ee
Expanding the exponential we find to second order in $k$:
\be\label{PP}
-i\omega\vek P^{\text{vy}} = -i\omega\vek P -\omega\vek k\cdot\vek Q -i\omega\vek R - \frac{1}{V}\int_V (\vek k\cdot\vek j)(\vek k\cdot\vek r)\vek r\diff^3r.
\ee
Furthermore, we obtain
\be\label{MM}
i\vek k\times\vek M^{\text{vy}} = i\vek k\times\vek M 
+ \frac{\vek k}{2V}\cdot\int_V (\vek j\vek r-\vek r\vek j)(\vek k\cdot\vek r)\diff^3r,
\ee
and
\be\label{QQ}
\omega\vek k\cdot\vek Q^{\text{vy}} = \omega\vek k\cdot\vek Q 
+ \frac{\vek k}{V}\cdot\int_V (\vek j\vek r+\vek r\vek j)(\vek k\cdot\vek r)\diff^3r,
\ee
Eqs. \eqref{PP}-\eqref{QQ} show the relation between the ``dipole'' and ``quadrupole'' terms in \eqref{Vinexp} compared to the usual ones. For example, \eqref{MM} shows that the difference $i\vek k\times(\vek M^{\text{vy}}-\vek M)$ is given by a magnetic quadrupole term.

Summing the contributions to the induced current, we obtain
\begin{align}\label{conn10}
&-i\omega\vek P^{\text{vy}}+i\vek k\times\vek M^{\text{vy}}+\omega\vek k\cdot\vek Q^{\text{vy}}/2  \\
= &-i\omega\vek P+i\vek k\times\vek M-\omega\vek k\cdot\vek Q/2 - i\omega\vek R. \nonumber
\end{align}
Eq. \eqref{conn10} could have been found directly by comparing \eqref{eq:Multipole} and \eqref{Vinexp}.

One may think that $\vek M^{\text{vy}}$ and $\vek M$, and the corresponding permeabilities, are equal in the limit $ka \to 0$, since then the $\exp(-i\vek k\cdot\vek r)$ factor in the integrand in \eqref{PMQM} tends to unity. Surprisingly, this is however not true in general. As an example, consider a metamaterial with a center of symmetry in the unit cell, which is taken as the origin. We must have
\be\label{symas}
\vek j(-\vek r)=\vek j(\vek r) \quad\text{when }\vek k\to 0,
\ee
which means that $\vek M\to 0$ as $\vek k\to 0$. In other words, $\vek M = \bigO(k)$. This can also be realized from Faraday's law: When there is no magnetoelectric coupling, $\vek M$ is proportional to $\vek B$, i.e., $\vek M = \vek\chi\vek B = \vek\chi(\vek k\times\vek E)/\omega = \bigO(k)$ for some tensor $\vek\chi$. By expanding the exponential in the definition of $\vek M^{\text{vy}}$ \eqref{PMQM}, the connection between $\vek M^{\text{vy}}$ and $\vek M$ can be written
\be\label{MM0conn}
\vek M^{\text{vy}} = \vek M + \frac{-i}{2V}\int_V (\vek k\cdot\vek r) \vek r\times\vek j(\vek r)\diff^3r.
\ee
The factor $\vek k\cdot\vek r$ in the integrand destroys the odd inversion symmetry, so the integral does not vanish in general. Thus the integral is $\Theta(k)$, and may be equally important as $\vek M$ in the limit $ka\to 0$. Recall that the permeabilities are found from the $\bigO(k)$ part of $\vek M^{\text{vy}}$ and $\vek M$, respectively. In other words, even though both $\vek M^{\text{vy}}$ and $\vek M$ tend to zero, the permeabilities derived from $\vek M^{\text{vy}}$ and $\vek M$ may be different. The difference between the permeabilities will be explored numerically in Sec. \ref{sec:numdisc}.

Finally we note that in general, the quantities $\vek P^\text{vy}$, $\vek M^\text{vy}$, and $\vek Q^\text{vy}$ are dependent on the choice of origin inside the cell $V$. Since $\vek M^\text{vy}$ may be origin-dependent, so is the resulting permeability $\vek\mu_\text{vy}$. From the definition \eqref{PMQM} it follows that the relative size of the origin dependence of $\vek M^{\text{vy}}$ is negligible when $\omega a P^{\text{ll}} \ll M^{\text{vy}}$. Numerically, the origin dependence of $\vek\mu_\text{vy}$ turns out to be minor, as discussed in Sec. \ref{sec:numdisc}.

\subsection{Transversal -- longitudinal (tl) decomposition}\label{sec:TL}
Starting from the Landau--Lifshitz permittivity, it is natural to use a strategy to put ``as much as possible'' of the $\vek k$-dependent induced current into the magnetization, and therefore the permeability. The resulting permeability is a generalization of that in Chapt. XII of Landau and Lifshitz' textbook \cite{landau_lifshitz_edcm}, and in Silveirinha \cite{silveirinha07}. 

The induced current can be divided into two parts:
\be\label{translong}
\vek J = -i\omega\vek P^{\text{tl}} + i\vek k\times\vek M^{\text{tl}}.
\ee
In \eqref{translong} the part which is independent of $\vek k$ is put into the first term $-i\omega\vek P^{\text{tl}}$. Moreover, the $\vek k$-dependent part is divided into a longitudinal part (which is parallel to $\vek k$), and a transversal part. The longitudinal part is also absorbed by the $-i\omega\vek P^{\text{tl}}$ term, while the transversal part is taken care of by the magnetization term $i\vek k\times\vek M^{\text{tl}}$. In a coordinate system oriented such that $\vek k=k\vekh x$, we can write
\be\label{translongkx}
\vek J = (-i\omega P^{\text{tl}}_1, -i\omega P^{\text{tl}}_2 -ik M^{\text{tl}}_3, -i\omega P^{\text{tl}}_3 + ik M^{\text{tl}}_2),
\ee
where $P^{\text{tl}}_2$ and $P^{\text{tl}}_3$ are independent of $\vek k$. As in Subsec. \ref{sec:multipole} \eqref{eq:MExp2}, we express
\begin{align}
& M^{\text{tl}}_{i} = \omega\zeta^{\text{tl}}_{ij}E_j + k\nu^{\text{tl}}_{i1j}E_j/(\mu_0\omega) \nonumber\\
 &= \omega\zeta^{\text{tl}}_{ij}E_j + \frac{k\nu^{\text{tl}}_{i11}E_1}{\mu_0\omega} + \mu_0^{-1} (1-\vek\mu_{\text{tl}}^{-1})_{ij}B_j \nonumber\\
      &= \omega\zeta^{\text{tl}}_{ij}E_j + \frac{k\nu^{\text{tl}}_{i11}E_1}{\mu_0\omega} + \frac{1}{\mu_0\omega} \left[(1-\vek\mu_{\text{tl}}^{-1})\vek k\times\vek E\right]_i \label{MtlkE}
\end{align}
for some $\zeta^{\text{tl}}_{ij}$, $\nu^{\text{tl}}_{ilk}$, and $\vek\mu_{\text{tl}}$.

The induced current density can also be expressed
\begin{align}
J_i &= -i\omega\epsilon_0[\epsilon_{ij}(\omega,\vek k)-\delta_{ij}]E_j, \label{JLLeps} \\
    &= -i\omega\epsilon_0[\chi_{ij} + \alpha_{ikj}k_k/\epsilon_0 + \beta_{iklj}k_kk_l c^2/\omega^2]E_j, \label{JLLeps2}
\end{align}
where we have substituted the Landau--Lifshitz permittivity \eqref{LLepsgamma2}. Equating the $\bigO(k^2)$ part of \eqref{translongkx} and the last term in \eqref{JLLeps2}, we obtain
\be\label{mugammaTL}
1-\vek\mu_{\text{tl}}^{-1} = \matrise{ \beta_{3113} & -\beta_{3112} \\ -\beta_{2113} & \beta_{2112} }.
\ee
In an arbitrary coordinate system, the tensor \eqref{mugammaTL} can be written
\be
\left[1-\vek\mu_{\text{tl}}^{-1}\right]_{mn} = \varepsilon_{mip}\varepsilon_{njq}\frac{k_kk_lk_pk_q}{k^4}\beta_{iklj}.
\ee
For strongly spatially dispersive media, we have let the last term in \eqref{JLLeps2} contain the remainder ($\Theta(k^2)$ and higher order). Then $\beta_{iklj}$ and the resulting $\vek\mu_\text{tl}$ become dependent on $\vek k$.

The symmetry \eqref{receps} means, according to \eqref{LLepsgamma2}, that $\beta_{iklj}(\vek k)=\beta_{jkli}(-\vek k)$. This means that
\be
\vek\mu_\text{tl}^\text{T}(-\vek k) = \vek\mu_\text{tl}(\vek k).
\ee
In particular, if we only consider terms of $\vek\epsilon(\omega,\vek k)$ up to second order in $k$ (weakly spatially dispersive media), we have $\vek\mu_\text{tl}^\text{T} = \vek\mu_\text{tl}$.

As for the asymptotic behavior of $\vek\mu_\text{tl}$ as $\omega\to\infty$, recall that the microscopic field tends to a plane wave in this limit, approximately unaffected by the structure. Using \eqref{JmcJ} and \eqref{microje}, we find 
\be
\vek J=\frac{-i\omega\epsilon_0\vek E}{V}\int_V[\varepsilon(\vek r)-1]\diff^3r + \Delta\vek J,
\ee
where 
\be
\Delta\vek J = -\frac{i\omega\epsilon_0}{V}\int_V [\varepsilon(\vek r)-1]\vek f(\vek r)\e{-i\vek k\cdot\vek r}\diff^3r.
\ee
The asymptotic behavior of $\varepsilon(\vek r)$ as $\omega\to\infty$ is of the form \eqref{varepsas}. From \eqref{fasympt} it is clear that $\Delta\vek J\to 0$ faster than $\vek J$. By comparison to \eqref{JLLeps} the resulting Landau--Lifshitz permittivity becomes
\be\label{LLepsas}
\epsilon_{ij}(\omega,\vek k) = \frac{\delta_{ij}}{V} \int_V \varepsilon(\vek r)\diff^3r + F_{ij}(\omega,\vek k),
\ee
where $F_{ij}(\omega,\vek k)$ tends to zero faster than $\omega^{-2}$. The first term in \eqref{LLepsas} is independent of $\vek k$; thus it does not contribute to the $\Theta(k^2)$ term of the Landau--Lifshitz permittivity \eqref{LLepsgamma2}. The term $F_{ij}(\omega,\vek k)$ may contribute, but gives a $\beta_{iklj}$ that tends to zero as $\omega\to\infty$. In other words,
\be\label{asympttl}
\vek\mu_{\text{tl}}^{-1} \to \vek I \text{ when } \omega\to\infty.
\ee

Since $\beta_{iklj}$ are the second order coefficients of $\epsilon_{ij}(\omega,\vek k)$, we can apply the general result in Appendix \ref{sec:analtensor} to deduce that $\beta_{iklj}$ and therefore $\vek\mu_{\text{tl}}^{-1}$ are analytic in the upper half-plane. With \eqref{asympttl} we conclude that $\vek\mu_{\text{tl}}^{-1}$ is causal.

The relation between the permeability resulting from the magnetic moment density (Subsec. \ref{sec:multipole}) and that in \eqref{mugammaTL} can be found by subtracting \eqref{mugammaTL} and \eqref{munu}:
\be\label{relmumutl}
\vek\mu^{-1} - \vek\mu_{\text{tl}}^{-1} = \matrise{ (\gamma+\psi+\eta)_{3113} & -(\gamma+\psi+\eta)_{3112} \\ -(\gamma+\psi+\eta)_{2113} & (\gamma+\psi+\eta)_{2112} }.
\ee
In other words the difference is due to the electric quadrupole, magnetic quadrupole + electric octupole, and $\Theta(k^2)$ part of electric dipole. The difference $\vek\mu_{\text{vy}}^{-1} - \vek\mu_{\text{tl}}^{-1}$ can be expressed similarly as in \eqref{relmumutl}, however without the $\psi$ tensor.

We have chosen, somewhat arbitrarily, to associate the entire $\Theta(k)$ term of the transversal current with the magnetization $\vek M^{\text{tl}}$. The $\Theta(k)$ term could be associated with polarization $\vek P^{\text{tl}}$ instead, or shared between the two. This has however no influence on the permeability \eqref{mugammaTL}, being defined from the $\bigO(k^2)$ term.

Since the permeability $\vek\mu_\text{tl}$ is derived from the Landau--Lifshitz total permittivity $\vek\epsilon(\omega,\vek k)$, which in turn is found from $\vek J$ and $\vek E$ with \eqref{EmcE} and \eqref{JmcJ}, it follows that $\vek\mu_\text{tl}$ is not dependent on the choice of origin.

\subsection{Other decompositions}\label{sec:other}
Clearly there are infinite number of ways to decompose the induced current, obtaining ``$\vek P$'', ``$\vek M$'', and possibly other ``multipole'' terms. The possible decompositions fall roughly into two categories. In the first category the magnetization vector is defined from an integral of the microscopic current. Examples include \eqref{eq:MultipoleVectorsMagn} and \eqref{PMQM}. The analyticity of the resulting inverse permeabilities, asymptotic behavior, and connection to the Landau--Lifshitz permittivity follow in the same way as in Subsec. \ref{sec:multipole}. In the second category the magnetization is defined from a certain division of the $\bigO(k^2)$ part of the induced current, by including any desired part of the $\beta_{iklj}$ tensor in \eqref{LLepsgamma2}. Then the properties of the resulting $\vek\mu^{-1}$ can be explored along the lines in Subsec. \ref{sec:TL}. Of course, not all such definitions lead to an analytic $\vek\mu^{-1}$; this must be ensured by carefully considering the frequency dependence of the division. Also, to ensure that $\vek\mu^{-1}$ is a tensor, the division of $\beta_{iklj}$ must be possible to formulate in tensor form.

\section{Numerical results}\label{sec:numdisc}
We will now consider some concrete examples of 2d metamaterials, using a finite-difference-frequency-domain (FDFD) numerical method \cite{costa09,dirdal18}. The metamaterial unit cells, and the associated, inverse permeability element $33$ (perpendicular to the unit cell figures) are shown in Figs. \ref{fig:annulus}-\ref{fig:splitringU} for $k=0$. For all examples except that in Fig. \ref{fig:annulusdiel}, we have used silver inclusions described by a Drude--Lorentz model with parameters from Ref. \cite{rakic98}.

We observe that the different permeabilities are identical in the low frequency limit. However, for the dielectric inclusions (Fig. \ref{fig:annulusdiel}), the \emph{relative} differences are relatively large, and do not vanish in the low frequency limit. For $\omega a/c > 0.6$, corresponding to $a/\lambda>0.1$ ($\lambda$ is the vacuum wavelength), the differences between the permeabilities are quite visible for all examples except the split ring resonator medium (Fig. \ref{fig:splitring}).

Note that although the definition of $\vek\mu_\text{vy}$ is similar to that of $\vek\mu$, in the examples $\vek\mu_\text{vy}$ is closer to $\vek\mu_\text{tl}$ in magnitude.

In Fig. \ref{fig:splitringU} we observe the origin dependence of the permeabilities. The permeability $\vek\mu_\text{tl}$ is origin-independent by definition, while $\vek\mu$ and $\vek\mu_\text{vy}$ are dependent on the choice of origin. The origin dependence is however rather weak in the considered frequency range. In general the origin dependence of $\vek\mu_\text{vy}$ seems to be weaker than that of $\vek\mu$. In fact, for the examples in Fig. \ref{fig:annulus}-\ref{fig:twobars} the origin dependence of $\vek\mu_\text{vy}$ turned out to be negligible (not shown).

In Fig. \ref{fig:splitringU} we find that for larger frequencies, the imaginary parts of the three permeabilities can be negative. Clearly the medium response must be highly nonlocal in this region; in the presence of spatial dispersion the condition for passivity is formulated in terms of the Landau--Lifshitz permittivity $\vek\epsilon(\omega,\vek k)$ (see \eqref{passivityreal}).

The causal properties of the inverse permeabilities $\vek\mu^{-1}$, $\vek\mu_\text{vy}^{-1}$, and $\vek\mu_\text{tl}^{-1}$, proven in Sec. \ref{sec:indcurr}, have been verified numerically for the metamaterials in Fig. \ref{fig:annulusP}-\ref{fig:splitringU}a using a Lorentzian model for the microscopic permittivity. This is done by first computing the (3,3) elements of the inverse permeabilities over a large bandwidth (such that the asymptotic limit can be seen). Then the results are Fourier transformed, and verified to be vanishing small for negative time.

Although the inverse permeabilities are causal, the permeabilities are generally not. This was noted for the $\vek\mu_\text{vy}$ permeability in Ref. \cite{Yaghjian2014}. Note that the inverse permeability is the natural response quantity appearing when expressing $\vek M$ from the fundamental field $\vek B$ (or expressing $\vek M$ from the applied current density $\vek J_\text{ext}$, using \eqref{eq:MExpn}, \eqref{EJLL}, and \eqref{munu}). Therefore, the inverse permeability is causal. Proving that the permeability itself is causal, from the causality of the inverse permeability, is possible only in certain special cases \cite{landau_lifshitz_edcm,Yaghjian2014}. For example, when the inverse permeability is scalar, and $\im\mu^{-1}$ takes only negative values, the inverse permeability turns out to be zero-free in the upper half-plane $\im\omega>0$. Then the permeability becomes causal. Otherwise, as for the metamaterials in Fig. \ref{fig:annulusP}-\ref{fig:splitringU}a, the permeabilities are noncausal despite the inverse permeabilities being causal.

%This is done by computing the (3,3) elements of the inverse permeabilities over a bandwidth $-50\pi c/a \leq\omega\leq 50\pi c/a$ using 30000 frequency points, and Fourier transforming the result (not shown). The microscopic permittivity was assumed Lorentzian with resonance frequency $5\pi c/a$, bandwidth $0.5\pi c/a$, and top value $i100$. The microscopic permittivity was assumed Lorentzian with resonance frequency $5\pi c/a$, bandwidth $0.5\pi c/a$, and top value $i100$.

\section{Discussion and Conclusion}\label{discconcl}
\bgroup
\def\arraystretch{1.3}
\begin{table*}[t]
	\begin{tabular}{| l | c | c | c | c |}
		\hline
		& \ref{sec:multipole} & \ref{sec:VY} & \ref{sec:TL} & \ref{sec:LL} \\
		& Multipole & Vinogradov--Yaghjian & Transversal--longitudinal & Landau--Lifshitz, $\vek\mu_\text{ll}=\vek I$ \\
		\hline
		Number of $\vek J$ expansion terms & $\infty$ ($\vek P$, $\vek M$, $\vek Q$, $\vek R$, \ldots) & 3 ($\vek P^\text{vy}$, $\vek M^\text{vy}$, $\vek Q^\text{vy}$) & 2 ($\vek P^\text{tl}$, $\vek M^\text{tl}$) & 1 ($\vek P^\text{ll}$) \\
		Causal, $\vek\mu^{-1}$ analytic for $\im\omega>0$& yes & yes & yes & $\vek G(\omega,\vek k)$ causal \\  
		Causal, $\vek\mu$ analytic for $\im\omega>0$& no (i.g.) & no (i.g.) & no (i.g.) & $\vek\epsilon(\omega,\vek k)$ causal \\  
		For $\omega\to\infty$ and fixed $\vek k$ & $\vek\mu\to\text{const}$ & $\vek\mu_\text{vy}\to\vek I$ & $\vek\mu_\text{tl}\to\vek I$ & $\vek\epsilon(\omega,\vek k)\to\vek I$ \\ 
		Sign of $\im\vek\mu$ (for diagonal $\vek\mu$) & both (i.g.) & both (i.g.) & both (i.g.) &  $\omega[\vek\epsilon(\omega,\vek k) - \vek\epsilon(\omega,\vek k)^\dagger] \text{ pos.}$ \\
		Symmetry & - & - & $\vek\mu_\text{tl}^\text{T}(-\vek k) = \vek\mu_\text{tl}(\vek k)$ & $\vek\epsilon^\text{T}(\omega,-\vek k) = \vek\epsilon(\omega,\vek k)$ \\ 
		Origin dependence & yes (i.g.) & yes (i.g.) & no & no \\ 
		\hline
	\end{tabular}
	\caption{General properties of induced current expansions and associated permeabilities; i.g. = in general. For the Landau--Lifshitz formulation the permeability is trivial, and the table column rather displays the properties of the permittivity tensor $\vek\epsilon(\omega,\vek k)$.}\label{tab:prop}
\end{table*}

In conclusion we have considered four definitions of permeability for periodic metamaterials, and their properties. The properties of the induced current decompositions and associated permeabilities are summed up in Table \ref{tab:prop}.

Having considered several different definitions of the magnetic permeability, it is natural to ask which one to prefer. Of course there is not a simple answer to this question. The Vinogradov--Yaghjian decomposition has the advantage of representing all induced current with only three terms. On the other hand, the conventional multipole decomposition has a clear physical interpretation; in particular the permeability $\vek\mu$ is induced from the magnetic moment density $\vek M$. However, the asymptotic behavior for $\omega\to\infty$ and fixed $\vek k$ is not necessarily $\vek\mu\to\vek I$, and the origin dependence is generally larger than that of $\vek\mu_\text{vy}$. The permeability $\vek\mu_\text{tl}$ has a less direct physical interpretation compared to $\vek\mu$, but has the nice properties that it is independent of the choice of origin, and symmetric. In addition it is appealing that it contains ``as much as possible'' of the $\bigO(k^2)$ part of the Landau--Lifshitz permittivity.

For weakly spatially dispersive media where the higher order $\bigO(k^3)$ terms are ignored, all permeabilities are independent of $k$. For $\vek\mu$ and $\vek\mu_\text{vy}$, higher order terms are included by allowing $\nu_{ilj}$ in \eqref{eq:MExp} and \eqref{constitutiveMVY} to be dependent on $\vek k$. For $\vek\mu_{\text{tl}}$, higher order terms are included by letting $\beta_{iklj}$ in \eqref{LLepsgamma2} be dependent on $\vek k$. In all these cases the highest order term in the Taylor expansions absorbs the remainder, making the permeabilities dependent on $\vek k$ in a straightforward way. For strongly spatially dispersive media, this could perhaps be useful in certain cases where the magnetization part of the induced current dominates.  

Despite the induced current being exactly represented by the expansion terms, neither of the permeabilities can alone describe the entire $\Theta(k^2)$ part of the Landau--Lifshitz permittivity. Therefore, even for weakly spatially dispersive media, we cannot always use one of the permeabilities in addition to a permittivity in Fresnel equations to describe reflection and transmission at an interface. When using the Fresnel equations, the errors will be dependent on the impact of the missed terms, but also induced by the fact that the conventional boundary conditions are not necessarily valid for the fundamental Floquet mode fields \cite{haagenvik18}. In the multipole expansion, the missed terms are the $\Theta(k^2)$ part of $\vek P$, $\vek Q$, and $\vek R$. In the Vinogradov--Yaghjian decomposition, the missed terms are the $\Theta(k^2)$ part of $\vek P^\text{vy}$ and $\vek Q^\text{vy}$. In the transversal--longitudinal decomposition the missed term is the $\Theta(k^2)$ part of $\vek P^\text{tl}$. Here we have assumed a nongyrotropic medium.

The semi-infinite case has been studied numerically in a separate work \cite{haagenvik18}. It was found that Fresnel equations with the three permeabilities in Subsecs. \ref{sec:multipole}-\ref{sec:TL} give accurate results for 2d metamaterials which mimic natural magnetism, in a frequency range with nontrivial magnetic response. The frequency range where the prediction of Fresnel's equation is accurate, is where the three permeabilities are approximately equal. Considering the numerical examples in Sec. \ref{sec:numdisc}, we can therefore expect that the permeabilities (except the trivial one in Subsec. \ref{sec:LL}) are useful in Fresnel's equation in the range where they approximately coincide.

For media with strong electric quadrupole response, and/or higher order multipoles, the basic Fresnel equation will not give an accurate prediction. The permeability can still be relevant, provided additional boundary conditions for the particular structure are found \cite{Golubkov1995,raab13,yaghjian14,Silveirinha14,yaghjian16}. In these cases, a better alternative could perhaps be to calculate the reflection and transmission using exact mode matching techniques, or even e.g. finite-difference-time-domain simulations.

It is natural to ask if the permeabilities are useless in the frequency ranges where they cannot be used to predict the reflection from a semi-infinite structure. Although the permeabilities have limited use in these cases, it is convenient to have definitions which are valid for all frequencies. This makes it possible to apply Kramers--Kronig relations and other theoretical constraints which are formulated for the entire frequency range. Although the permeabilities lose their usual physical interpretation for sufficiently large frequencies, they are still physical in the sense that they are found from the physical, microscopic fields using the particular definition. For example, $\vek\mu$ in Subsec. \ref{sec:multipole} results from a magnetization $\vek M$ which quantifies the magnetic moment of the unit cell.

\section{Acknowledgements}
The authors would like to thank Dr. Arthur Yaghjian for constructive feedback and fruitful discussions.

\begin{figure}[!ht]
	\center
	
	\subfloat[]{
		\label{fig:annulusP}\begin{tikzpicture} [scale=3.5]
		\path [draw=none,fill=gray, fill opacity = 1] (0,0) circle (0.45);
		\path [draw=none,fill=white, fill opacity = 1] (0,0) circle (0.30);
		\draw[-] (-0.5,-0.5)--(-0.5,0.5)--(0.5,0.5)--(0.5,-0.5)--(-0.5,-0.5);
		\node [] at (-0.22,0.30) {$\varepsilon$};
		\node [] at (-0.37,0.42) {$\varepsilon=1$};
		\draw[<->] (-0.5,-0.55)--(0.5,-0.55);
		\node [] at (0,-0.6) {$a$};
		\draw[->] (0,0)--(-0.30,0);
		\node [above] at (-0.14,0) {$0.3a$};
		\draw[->] (0,0)--(-0.3182,-0.3182);
		\node [right] at (-0.12,-0.15) {$0.45a$};
		\draw[->,very thick] (0,0)--(0.6,0);
		\node [right] at (0.6,0) {$\vek k$};
		\draw[->] (0.8,0)--(0.9,0);
		\node [right] at (0.9,0) {$\vekh x$};
		\draw[->] (0.8,0)--(0.8,0.1);
		\node [right] at (0.8,0.1) {$\vekh y$};
		\end{tikzpicture}}
	
	\subfloat[]{\label{fig:annulusdiel}
		\includegraphics[width=0.4\textwidth]{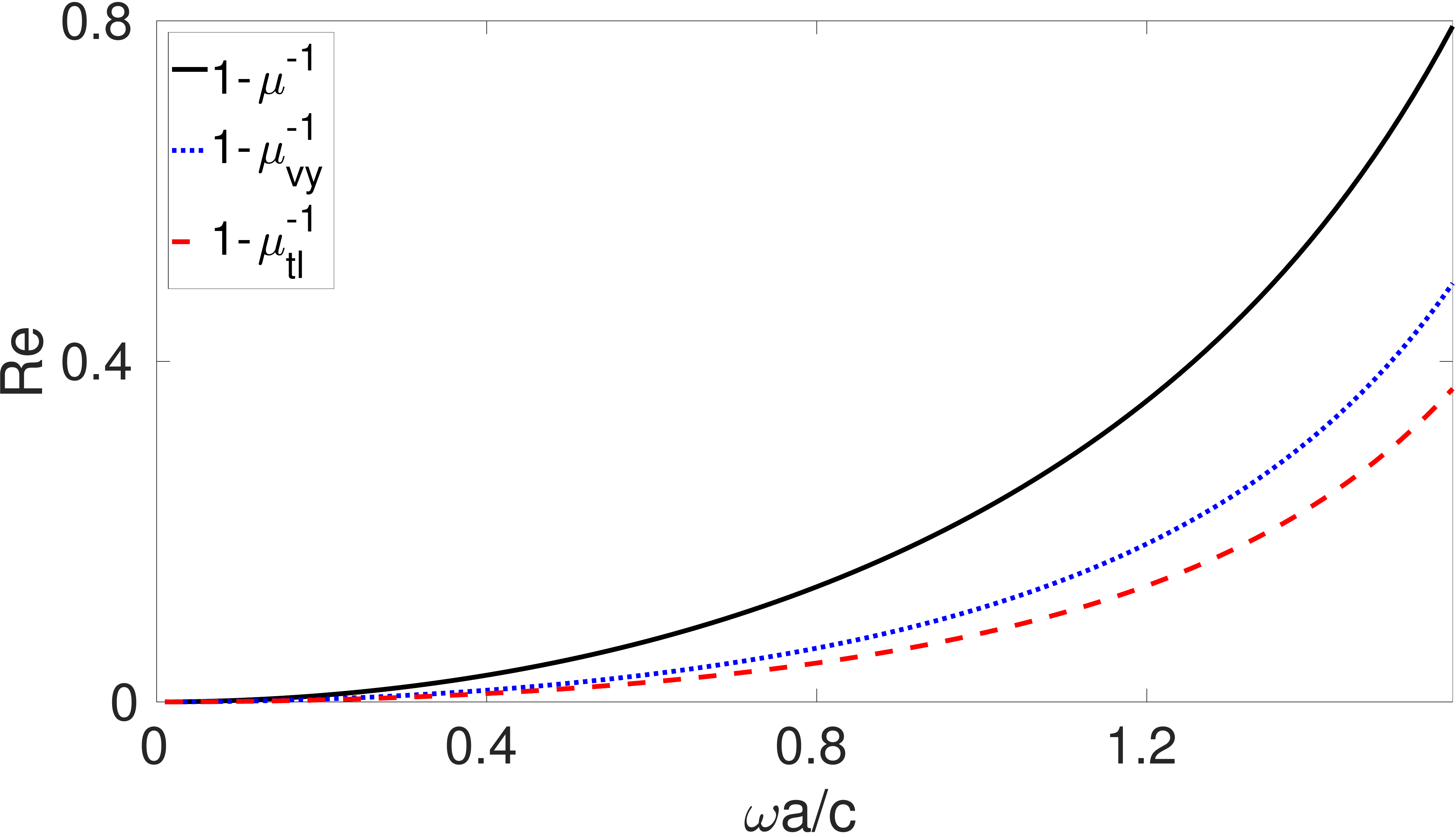}
	}
	
	\subfloat[]{\label{fig:annulusRe}
		\includegraphics[width=0.4\textwidth]{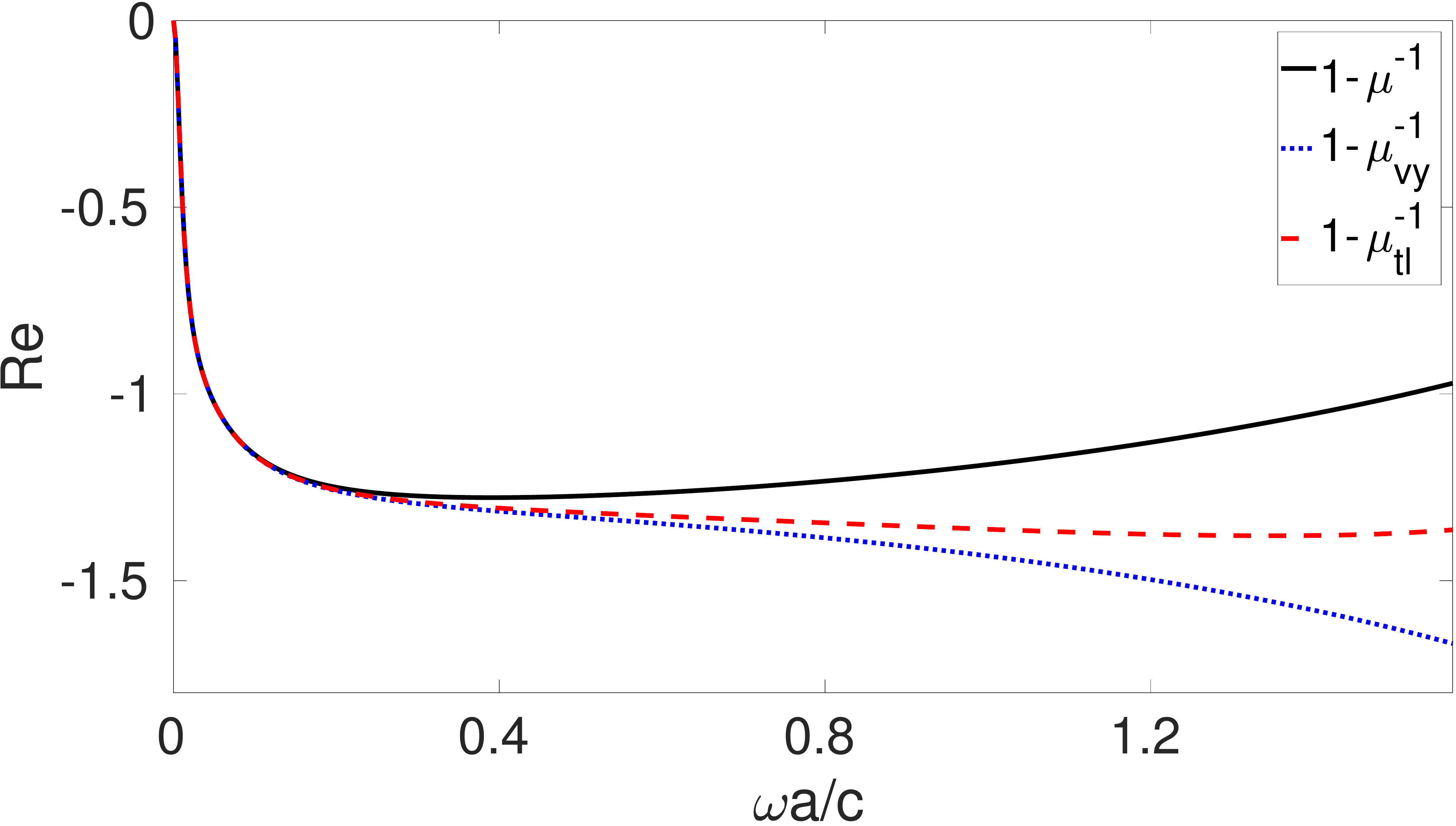}
	}
	
	\subfloat[]{\label{fig:annulusIm}
		\includegraphics[width=0.4\textwidth]{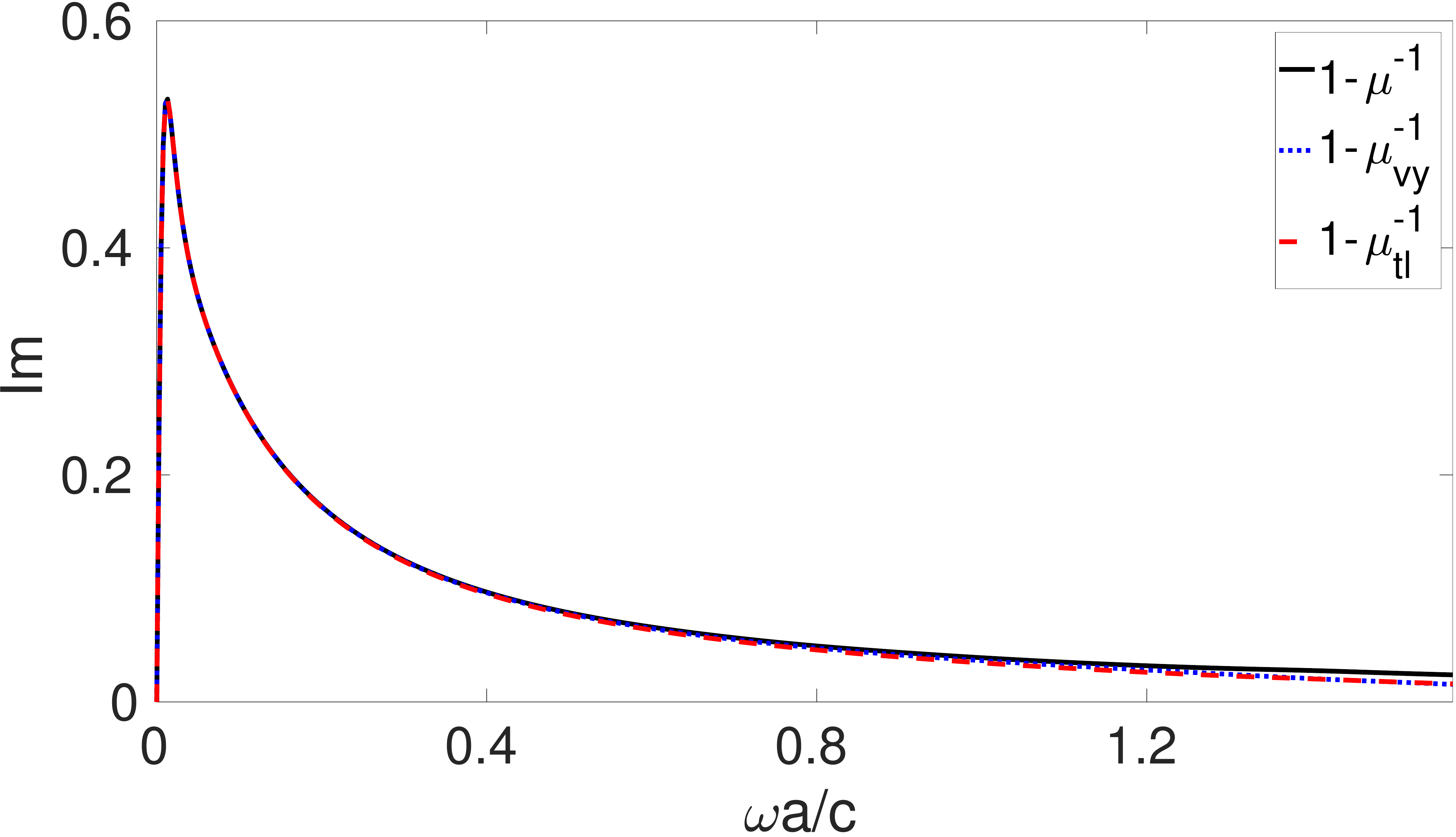}
	}
	
	\caption{(a) Unit cell with an annulus. (b) $1-\text{permeability}^{-1}$ when the annulus is a lossless dielectric ($\varepsilon=16$). Real (c) and imaginary (d) parts when the annulus is made from silver, and $a=1\,\mu$m.}
	\label{fig:annulus}
\end{figure}
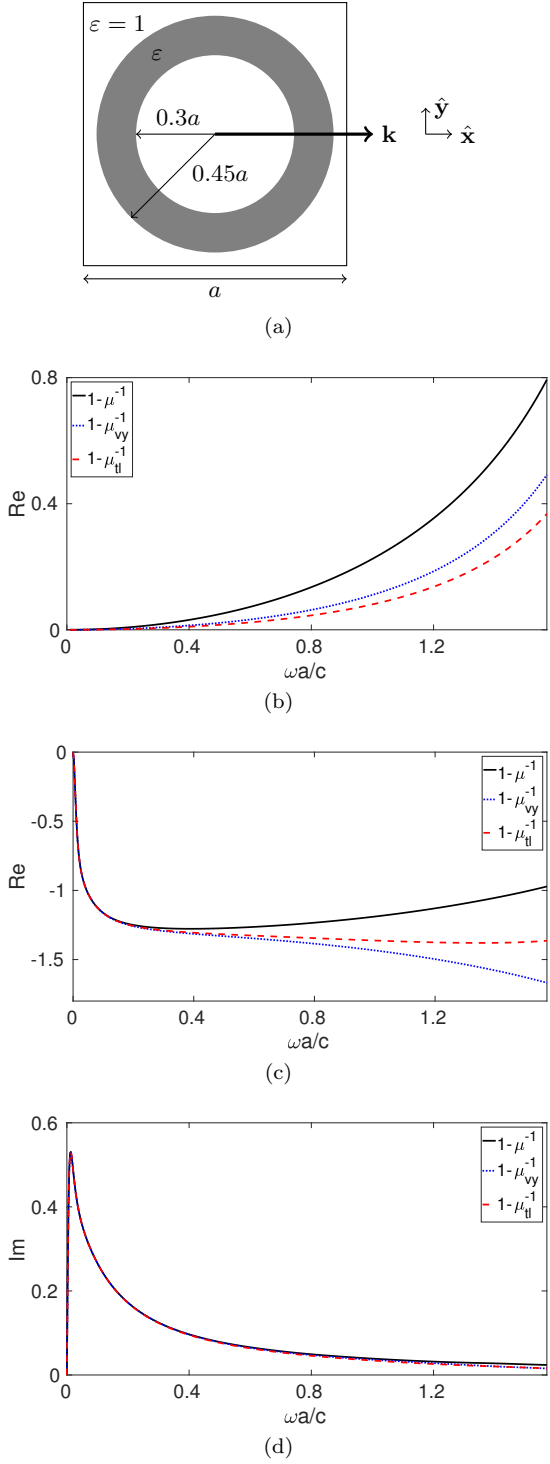

\begin{figure}[!tb]
	\center
	
	\subfloat[]{\label{fig:splitringP}\begin{tikzpicture} [scale=3.5]
		\path [draw=none,fill=gray, fill opacity = 1] (0,0) circle (0.45);
		\path [draw=none,fill=white, fill opacity = 1] (0,0) circle (0.30);
		\draw[-] (-0.5,-0.5)--(-0.5,0.5)--(0.5,0.5)--(0.5,-0.5)--(-0.5,-0.5);
		\node [] at (-0.22,0.30) {$\varepsilon$};
		\node [] at (-0.37,0.42) {$\varepsilon=1$};
		\draw[<->] (-0.5,-0.55)--(0.5,-0.55);
		\node [] at (0,-0.6) {$a$};
		\draw[->] (0,0)--(-0.30,0);
		\node [above] at (-0.14,0) {$0.3a$};
		\draw[->] (0,0)--(-0.3182,-0.3182);
		\node [right] at (-0.12,-0.15) {$0.45a$};
		\draw[->,very thick] (0,0)--(0.6,0);
		\node [right] at (0.6,0) {$\vek k$};
		\fill[black!0!white] (-0.01,-0.5) rectangle (0.01,-0.29);
		\fill[black!0!white] (-0.01,0.29) rectangle (0.01,0.5);
		\draw[->] (-0.05,0.27)--(-0.01,0.27);
		\draw[->] (0.05,0.27)--(0.01,0.27);
		\node [] at (0.02,0.21) {$0.02a$};
		\draw[->] (0.8,0)--(0.9,0);
		\node [right] at (0.9,0) {$\vekh x$};
		\draw[->] (0.8,0)--(0.8,0.1);
		\node [right] at (0.8,0.1) {$\vekh y$};
		\end{tikzpicture}}
	
	\subfloat[]{\label{fig:splitringRe}
		\includegraphics[width=0.4\textwidth]{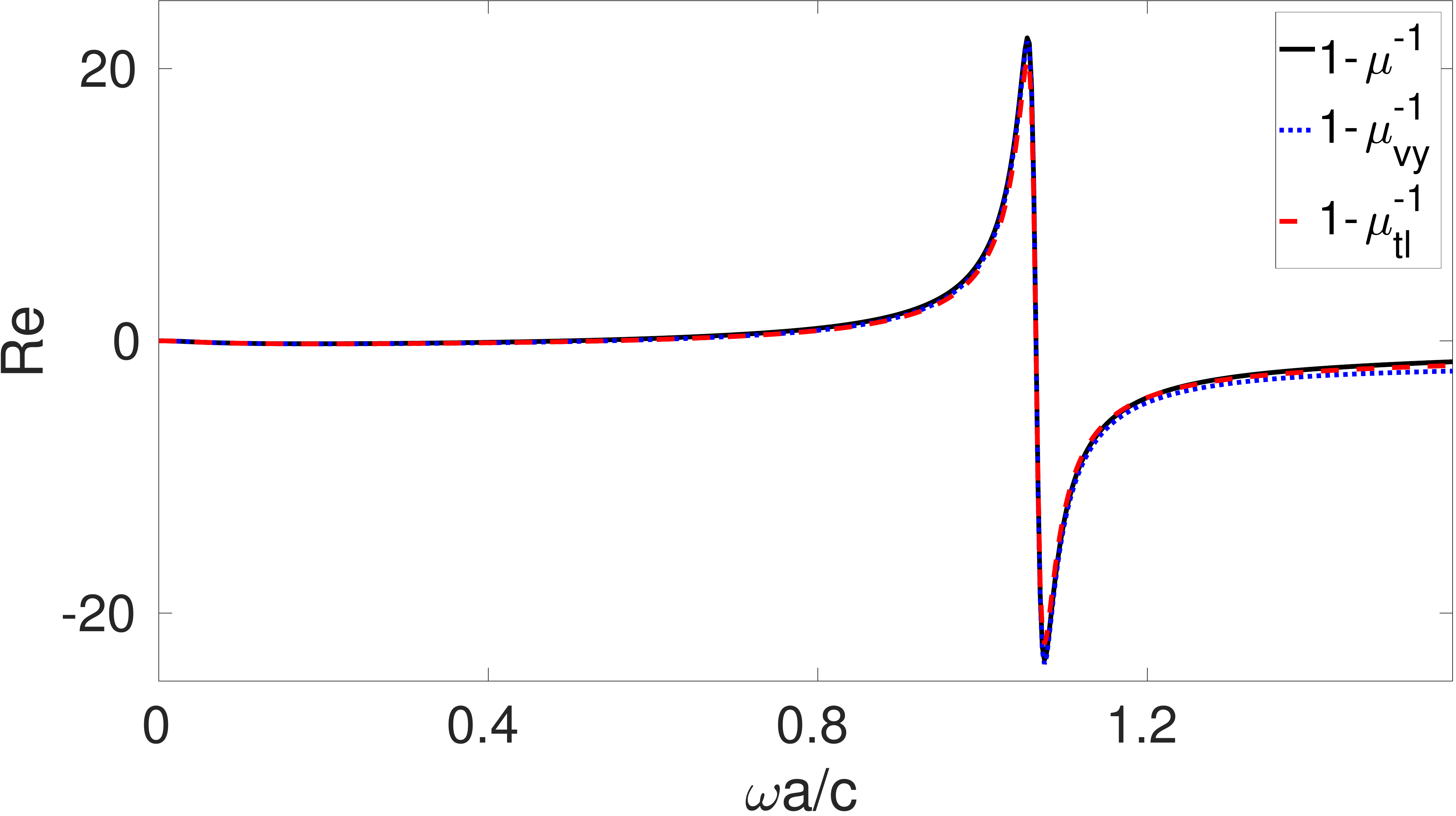}
	}
	
	\subfloat[]{\label{fig:splitringIm}
		\includegraphics[width=0.4\textwidth]{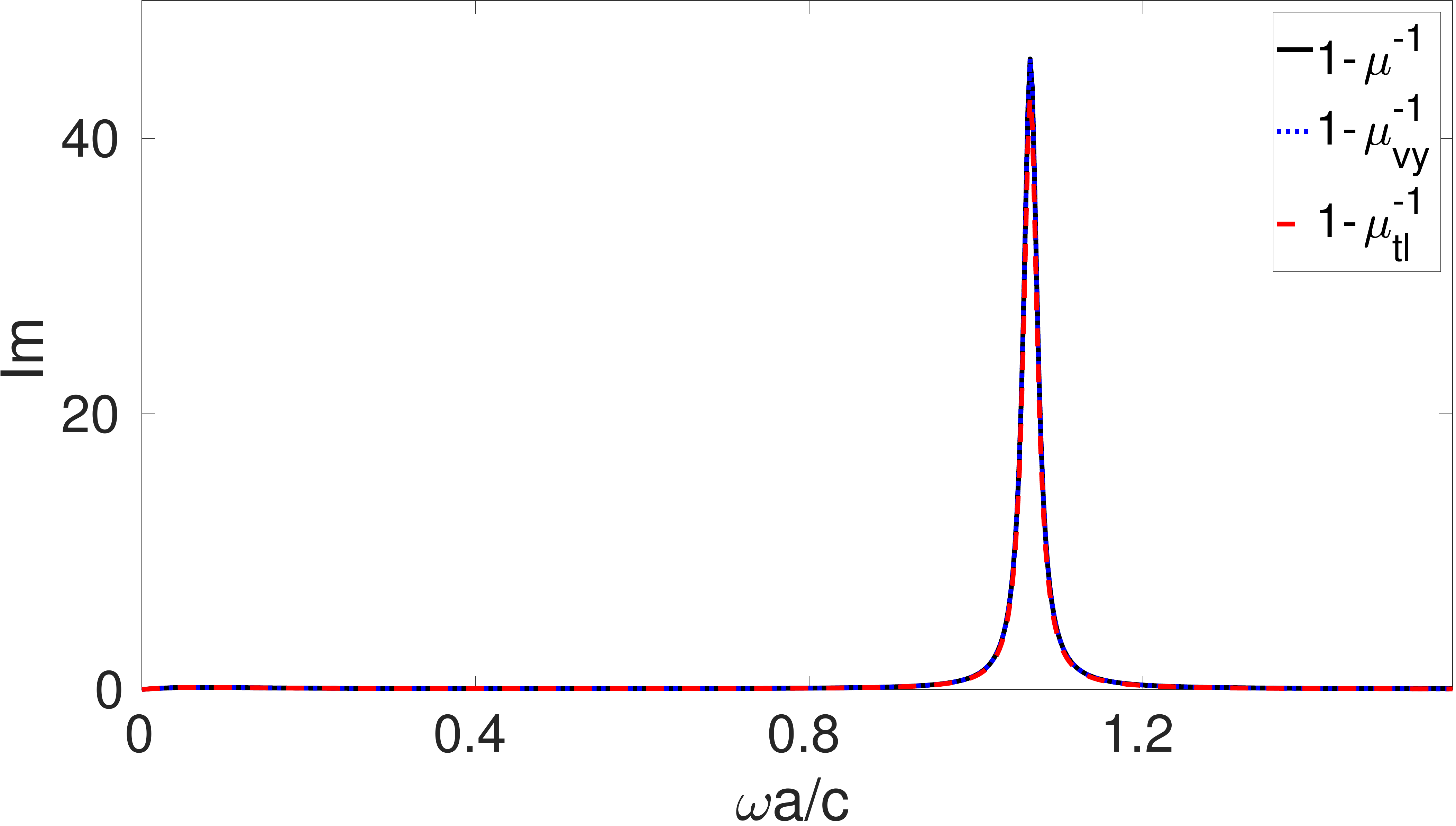}
	}
	
	\caption{(a) Unit cell with a split-ring resonator made from silver, $a=1\,\mu$m. Real (b) and imaginary (c) part of $1-\text{permeability}^{-1}$.}
	\label{fig:splitring}
\end{figure}
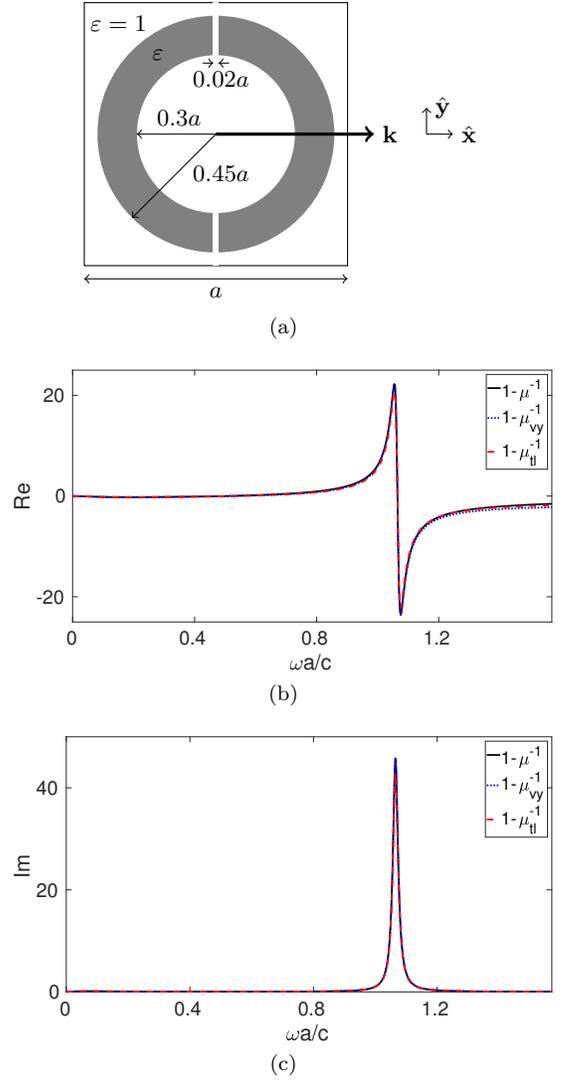

\cleardoublepage

\begin{figure}[!tb]
	\center
	
	\subfloat[]{\label{fig:twobarsP}
		\begin{tikzpicture} [scale=3.5]
		\fill[black!50!white] (0.15,-0.4) rectangle (0.35,0.4);
		\fill[black!50!white] (-0.35,-0.4) rectangle (-0.15,0.4);
		\draw[-] (-0.5,-0.5)--(-0.5,0.5)--(0.5,0.5)--(0.5,-0.5)--(-0.5,-0.5);
		\fill[black!0!white] (-0.52,-0.48) rectangle (0.52,-0.38);t
		\node [] at (0,0.3) {$\varepsilon=1$};
		\node [] at (-0.25,0.3) {$\varepsilon$};
		\node [] at (0.25,0.3) {$\varepsilon$};
		\draw[<->] (-0.15,-0.2)--(0.15,-0.2);
		\node [below] at (0,-0.2) {$0.3a$};
		\draw[<->] (-0.35,-0.44)--(-0.15,-0.44);
		\node [left] at (-0.335,-0.43) {$0.2a$};
		\draw[<->] (0.35,-0.44)--(0.15,-0.44);
		\node [right] at (0.335,-0.43) {$0.2a$};
		\draw[->,very thick] (0,0)--(0.6,0);
		\node [right] at (0.6,0) {$\vek k$};
		\draw[<->] (-0.5,-0.55)--(0.5,-0.55);
		\node [] at (0,-0.6) {$a$};
		\draw[<->] (0.25,0.4)--(0.25,0.5);
		\node [right] at (0.3,0.45) {$0.1a$};
		\draw[->] (0.8,0)--(0.9,0);
		\node [right] at (0.9,0) {$\vekh x$};
		\draw[->] (0.8,0)--(0.8,0.1);
		\node [right] at (0.8,0.1) {$\vekh y$};
		\fill[black!0!white] (-0.79,0) rectangle (-0.8,0.1);
		\end{tikzpicture}}
	
	\subfloat[]{\label{fig:twobarsRe}
		\includegraphics[width=0.4\textwidth]{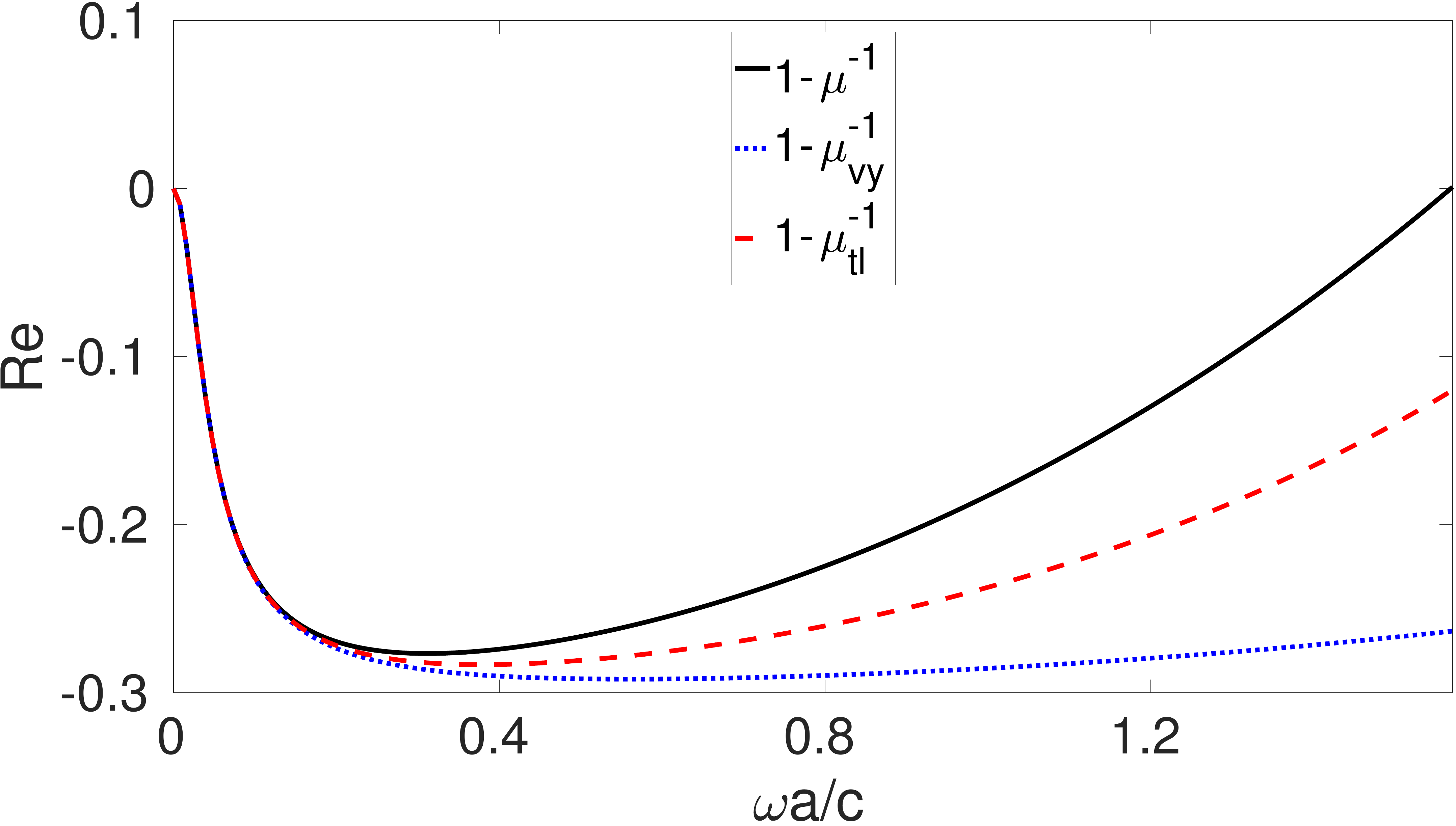}
	}
	
	\subfloat[]{\label{fig:twobarsIm}
		\includegraphics[width=0.4\textwidth]{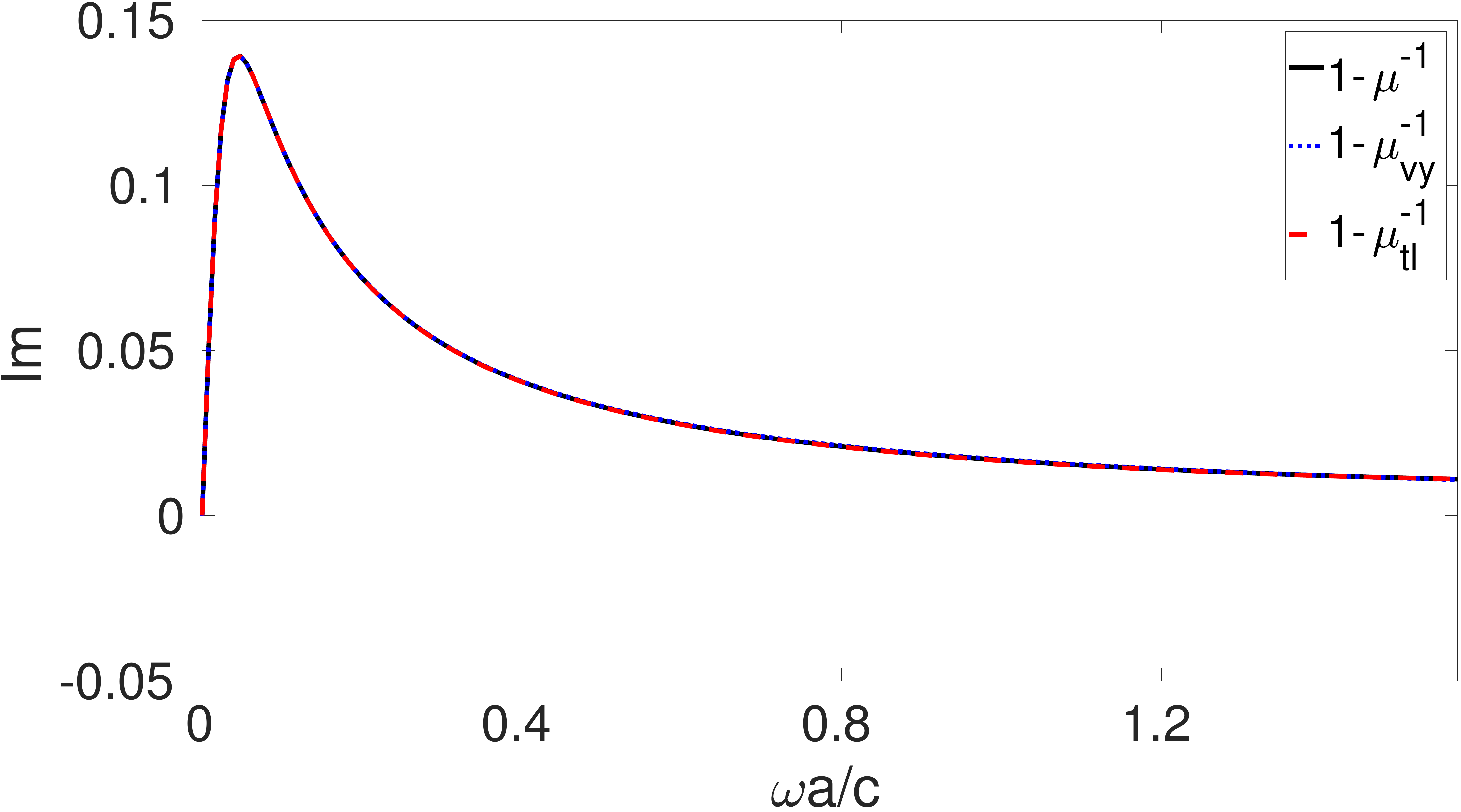}
	}
	
	\caption{(a) Unit cell with two bars made from silver, $a=1\,\mu$m. Real (b) and imaginary (c) part of $1-\text{permeability}^{-1}$.}
	\label{fig:twobars}
\end{figure}
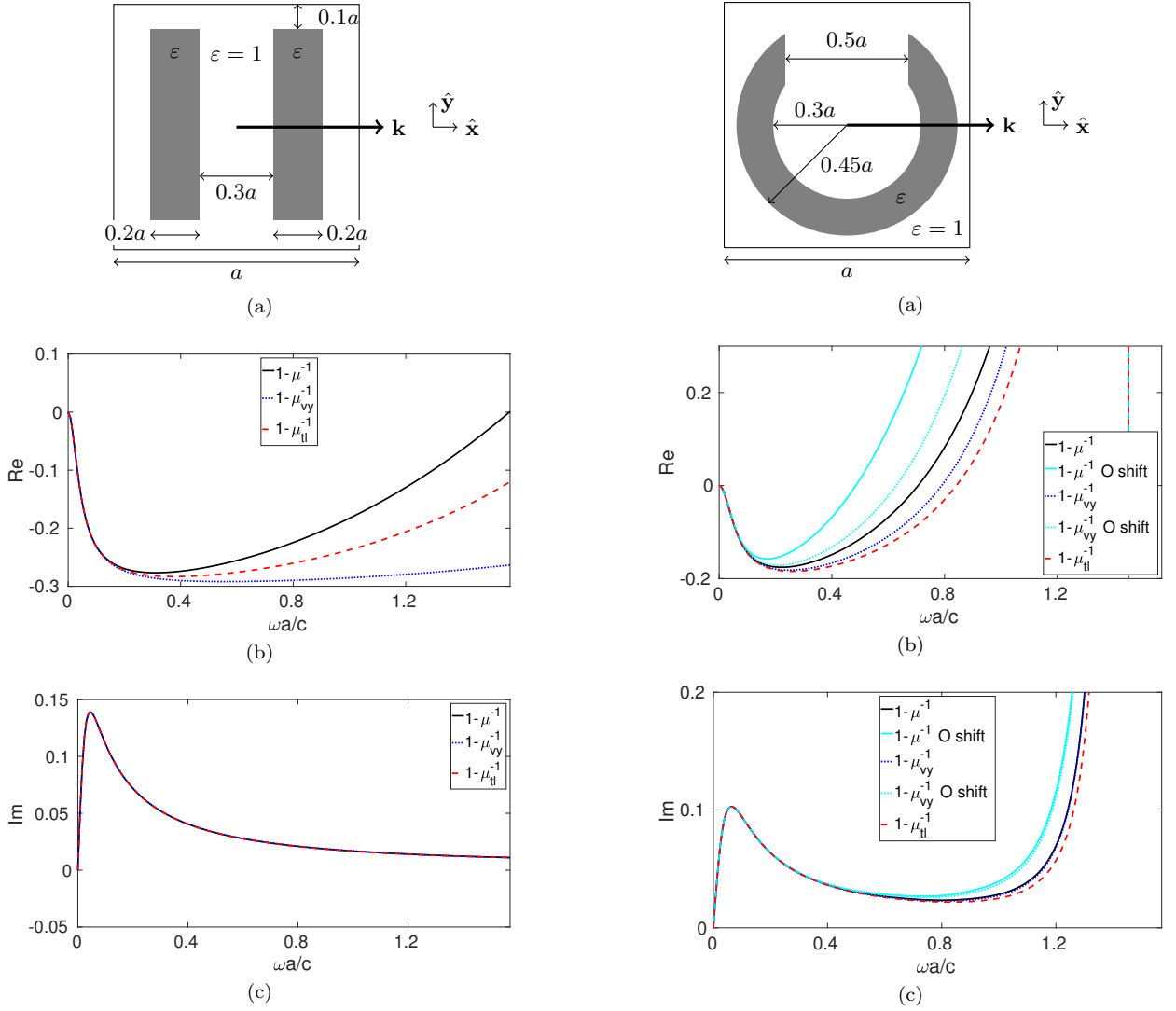

\begin{figure}
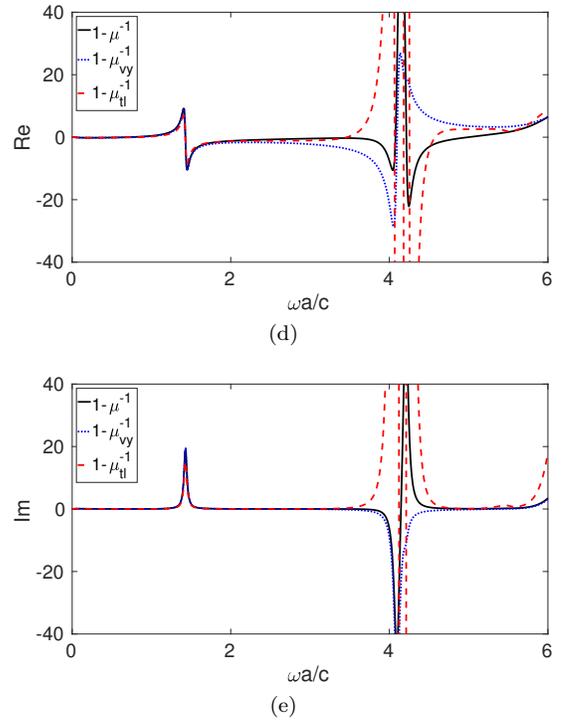

	\caption{(a) Unit cell with a ``U'' made from silver, $a=1\,\mu$m. Real (b) and imaginary (c) part of $1-\text{permeability}^{-1}$. Also shown are the results when the origin has been shifted from the center of the cell $(0,0)$ to top right corner $(a/2,a/2)$. In (d) and (e) the results are plotted for higher frequencies, demonstrating that the imaginary parts can have either sign in this region. This does not mean violation of passivity, but that the medium response is nonlocal.}
	\label{fig:splitringU}
\end{figure}

\begin{figure}[!tb]
	\center
	
	\subfloat[]{\label{fig:splitringUP}\begin{tikzpicture} [scale=3.5]
		\path [draw=none,fill=gray, fill opacity = 1] (0,0) circle (0.45);
		\path [draw=none,fill=white, fill opacity = 1] (0,0) circle (0.30);
		\draw[-] (-0.5,-0.5)--(-0.5,0.5)--(0.5,0.5)--(0.5,-0.5)--(-0.5,-0.5);
		\node [] at (0.22,-0.30) {$\varepsilon$};
		\node [] at (0.37,-0.42) {$\varepsilon=1$};
		\draw[<->] (-0.5,-0.55)--(0.5,-0.55);
		\node [] at (0,-0.6) {$a$};
		\draw[->] (0,0)--(-0.30,0);
		\node [above] at (-0.13,0) {$0.3a$};
		\draw[->] (0,0)--(-0.3182,-0.3182);
		\node [right] at (-0.14,-0.16) {$0.45a$};
		\fill[black!0!white] (-0.25,0.1) rectangle (0.25,0.45);
		\draw[->,very thick] (0,0)--(0.6,0);
		\node [right] at (0.6,0) {$\vek k$};
		\draw[<->] (-0.25,0.27)--(0.25,0.27);
		\node [] at (0,0.35) {$0.5a$};
		\draw[->] (0.8,0)--(0.9,0);
		\node [right] at (0.9,0) {$\vekh x$};
		\draw[->] (0.8,0)--(0.8,0.1);
		\node [right] at (0.8,0.1) {$\vekh y$};
		\end{tikzpicture}}
	
	\subfloat[]{\label{fig:splitringURe}
		\includegraphics[width=0.4\textwidth]{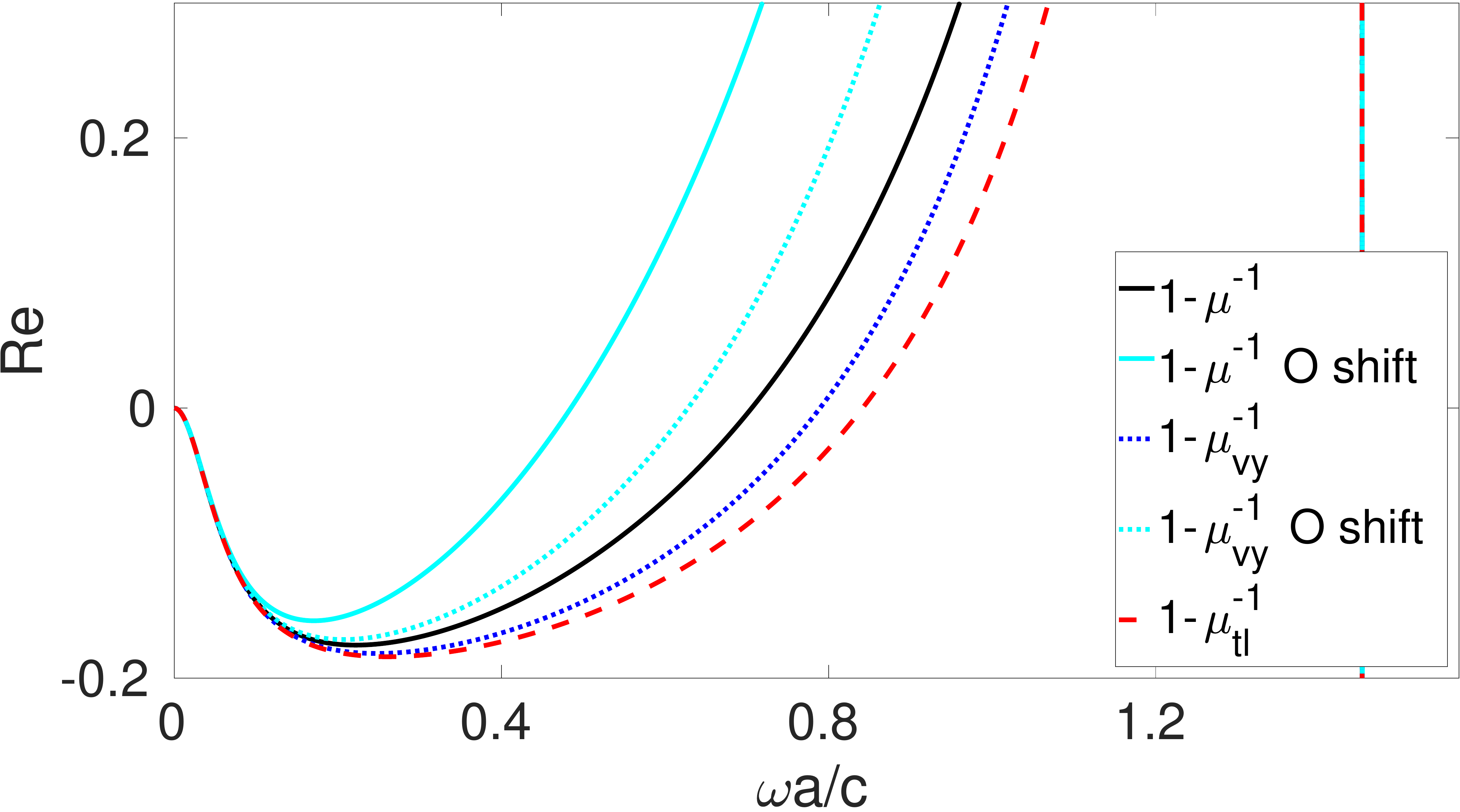}
	}
	
	\subfloat[]{\label{fig:splitringUIm}
		\includegraphics[width=0.4\textwidth]{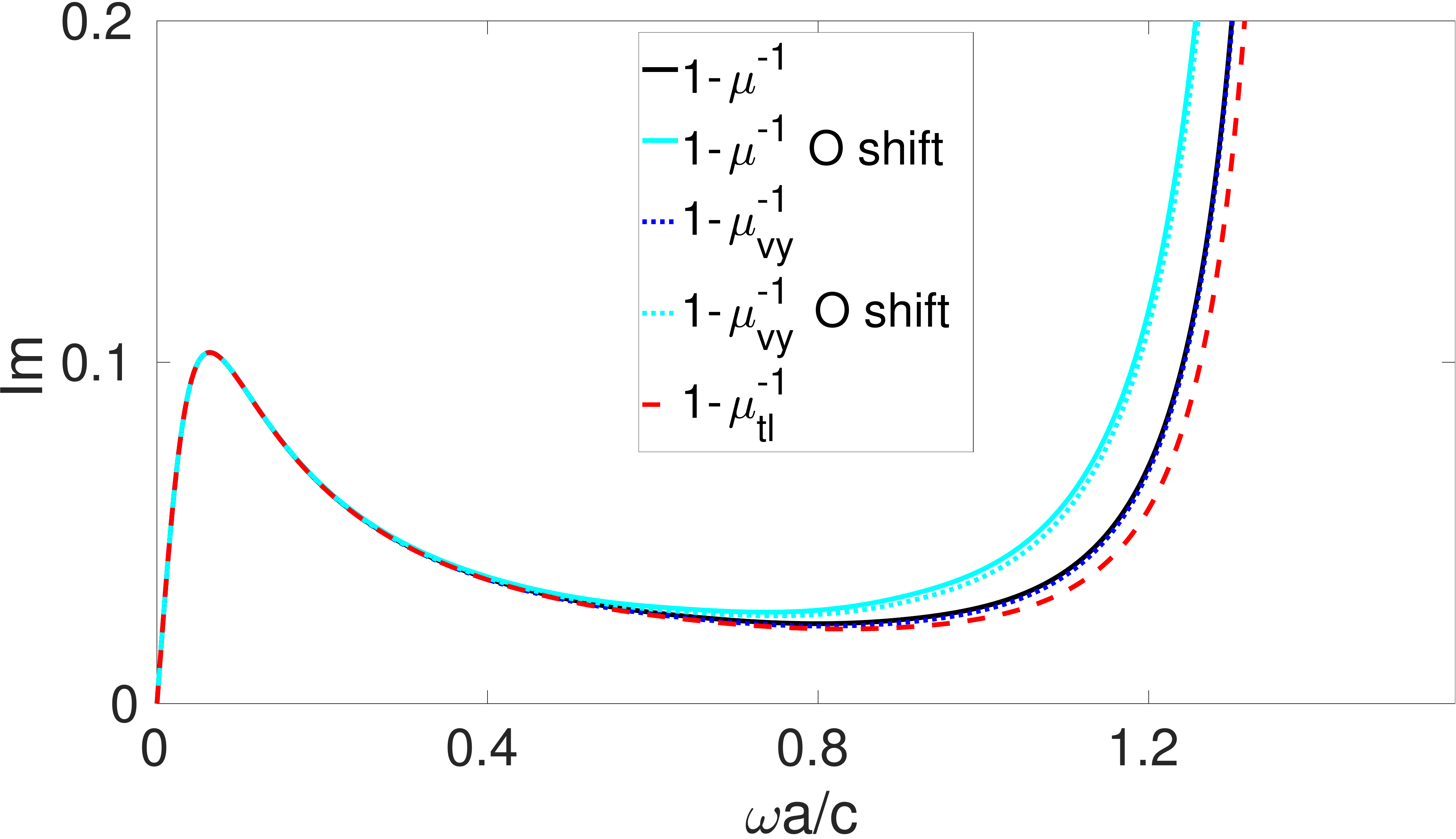}
	}
	
	\subfloat[]{\label{fig:splitringURee}
		\includegraphics[width=0.4\textwidth]{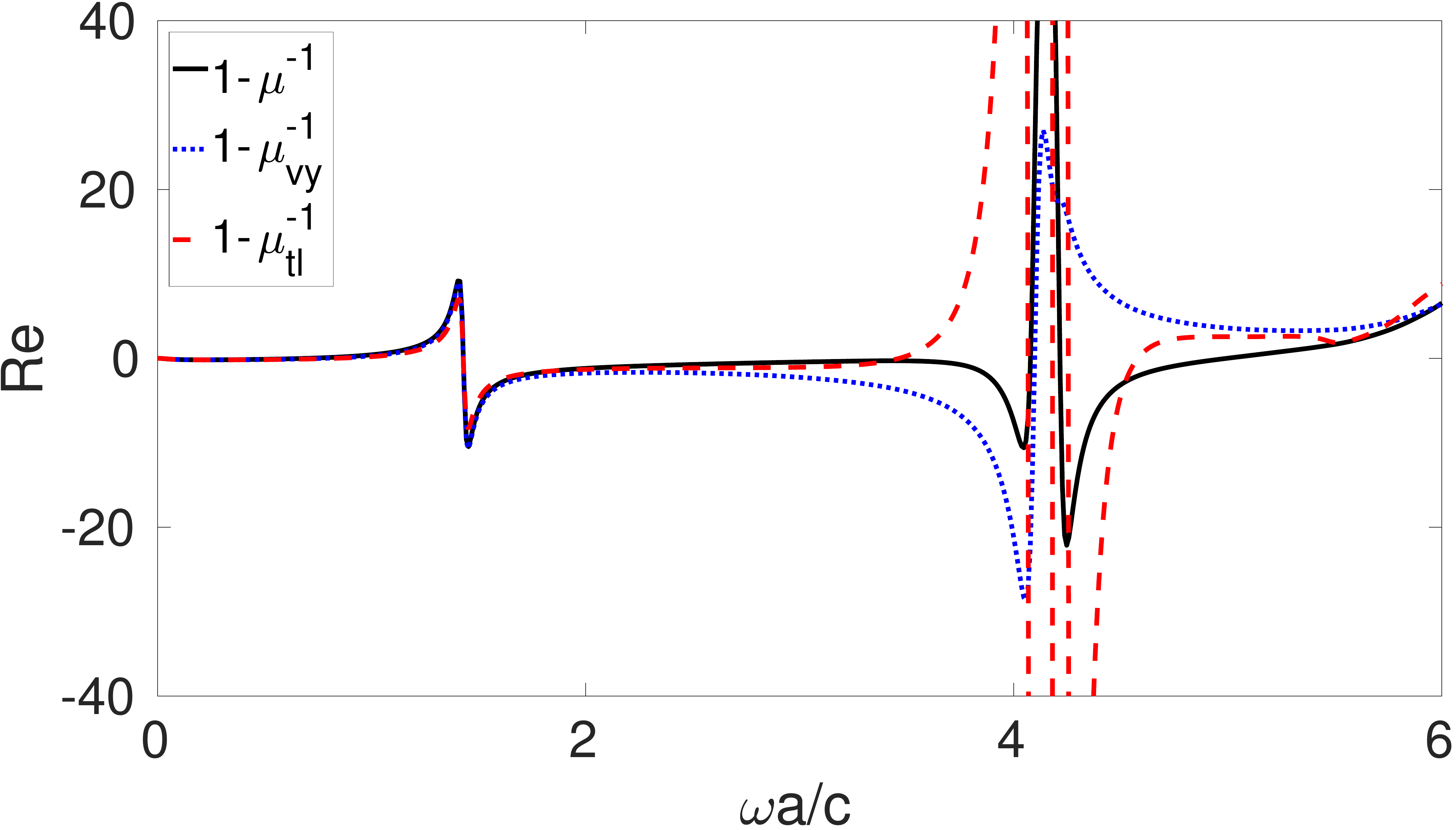}
	}
	
	\subfloat[]{\label{fig:splitringUIme}
		\includegraphics[width=0.4\textwidth]{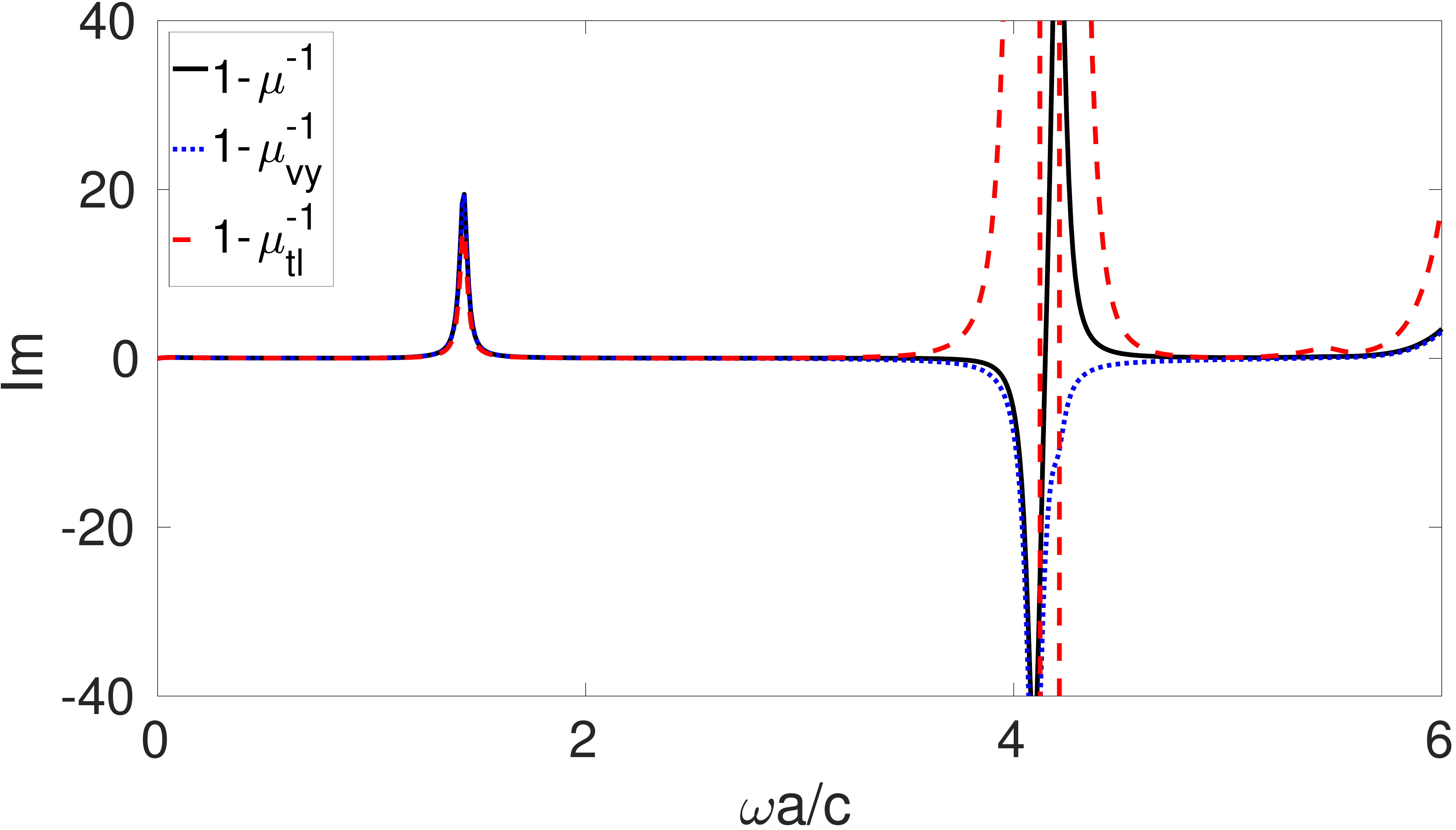}
	}
	
\end{figure}

\cleardoublepage

%\clearpage

\appendix

\section{Deriving $\vek k$-domain Maxwell equations for homogenized fields}\label{sec:maxwhom}
Our starting point is the microscopic Maxwell equations:
\begin{subequations}\label{mmaxwellmicro}
\begin{align}
\curl e &= i\omega\vek b, \\
\frac{1}{\mu_0}\curl b &= -i\omega\epsilon_0\vek e+\vek j(\vek r) + \vek J_\text{ext}\e{i\vek k\cdot\vek r}.
\end{align}
\end{subequations}
Since the structure is periodic, and the source is of the form $\vek J_\text{ext}\e{i\vek k\cdot\vek r}$ with constant $\vek J_\text{ext}$, all fields can be written in Floquet form. For example, $\vek e=\vek u_{\vek e}\e{i\vek k\cdot\vek r}$. Substituting into \eqref{mmaxwellmicro} we obtain
\begin{subequations}\label{mmaxwellmicro3}
\begin{align}
\nabla\times\vek u_{\vek e} + i\vek k\times\vek u_{\vek e} &= i\omega \vek u_{\vek b}, \\
\frac{1}{\mu_0}\nabla\times\vek u_{\vek b} + \frac{1}{\mu_0}i\vek k\times \vek u_{\vek b} &= -i\omega\epsilon_0\vek u_{\vek e} + \vek u_{\vek j} + \vek J_\text{ext}.
\end{align}
\end{subequations}
Recall that the periodic $\vek u$ functions can be written in terms of their Fourier components, as in \eqref{per}. 
Eqs. \eqref{mmaxwellmicro3} (and therefore \eqref{mmaxwellmicro}) are satisfied if and only if the Fourier components of \eqref{mmaxwellmicro3} satisfy:
\begin{subequations}\label{mmaxwellF}
\begin{align}
& i(\vek b_{lmn}+\vek k)\times\vek E_{lmn} = i\omega \vek B_{lmn}, \\
& \frac{1}{\mu_0}i(\vek b_{lmn}+\vek k)\times\vek B_{lmn} = -i\omega\epsilon_0\vek E_{lmn} + \vek J_{lmn},
\end{align}
\end{subequations}
for all $l,m,n$ except $l=m=n=0$, for which the set can be written  
\begin{subequations}\label{mmaxwellF0}
\begin{align}
i\vek k\times\vek E &= i\omega \vek B, \\
\frac{1}{\mu_0}i\vek k\times \vek B &= -i\omega\epsilon_0\vek E + \vek J + \vek J_\text{ext}.
\end{align}
\end{subequations}
Eqs. \eqref{mmaxwellF0} are the Maxwell equations for the fundamental Floquet modes, which we have taken to be the macroscopic fields. Eqs. \eqref{mmaxwellF} are the equations that the other Fourier components must satisfy.

The induced current $\vek J_{lmn}$ couples between sets with different indices. Defining $\sigma(\vek r)=-i\omega\epsilon_0[\varepsilon(\vek r)-1]$, we have
\begin{align}\label{JlmnE}
\vek J_{lmn} &= \frac{1}{V}\int \vek j(\vek r)\e{-i\vek k\cdot\vek r-i\vek b_{lmn}\cdot\vek r}\diff^3r \\
&= \frac{1}{V}\int \sigma(\vek r)\vek e(\vek r)\e{-i\vek k\cdot\vek r-i\vek b_{lmn}\cdot\vek r}\diff^3r \nonumber\\
&= \sum_{l'm'n'}\vek E_{l'm'n'} \cdot\frac{1}{V}\int \sigma(\vek r)\e{i(\vek b_{l'm'n'}-\vek b_{lmn})\cdot\vek r}\diff^3r \nonumber
\end{align}
By eliminating $\vek B_{lmn}$ from \eqref{mmaxwellF}-\eqref{mmaxwellF0}, and using \eqref{JlmnE}, we obtain a linear equation set in the form 
\be\label{lineqset}
\sum_n A_{mn} E_n=J_{\text{ext}}\delta_{m0},
\ee
where the matrix $A_{mn}$ depends on $\omega$, $\vek k$, and microscopic permittivity, but not the fields. The three indices $lmn$ and the index of the vector components have been combined into a single index $n$ or $m$, and the coordinate system is oriented such that $\vek J_\text{ext}$ is along one of the axes, corresponding to $m=0$. The elements $E_n$ of the new field vector contains the three components of each $\vek E_{lmn}$. From \eqref{lineqset} we note that all fields, e.g. $\vek E_{lmn}$ or $\vek E$, are proportional to $J_\text{ext}$.

\section{Macroscopic fields for arbitrary $ka$}\label{sec:work}
Here we will prove that the macroscopic fields (or fundamental Floquet mode fields) can be used to calculate the work done by the source in a unit cell, even for large wavenumbers. Consider first a source with a single, spatial Fourier component, $\vek j_\text{ext}(\vek r)=\vek J_\text{ext}(\vek k)\e{i\vek k\cdot\vek r}$. The work done by the source per unit volume and per unit time (after averaging over a period) is
\be\label{workdens}
p_\text{ext}=\frac{1}{2}\re\{\vek j_\text{ext}\cdot (-\vek e^*) \},
\ee
where $\vek e$ is the microscopic electric field. Substituting \eqref{spatJk} and \eqref{floqebe} we find
\be
p_\text{ext}=\frac{1}{2}\re\{\vek J_\text{ext}\cdot (-\vek u_{\vek e}^*) \},
\ee
which after averaging over a unit cell $V$ (using \eqref{EmcE}) becomes
\be\label{avworkE}
\langle p_\text{ext} \rangle=\frac{1}{2}\re\{\vek J_\text{ext}\cdot (-\vek E^*) \},
\ee
or
\be\label{avwork}
\langle p_\text{ext} \rangle=\frac{1}{2}\re\{\vek j_\text{ext}\cdot (-\mc E^*) \}.
\ee
In other words, we can find the work from the macroscopic field $\mc E$. 

For a source in the form
\be
\vek j_\text{ext}(\vek r) = \vek j_0(\vek r)\e{i\vek k_0\cdot\vek r},
\ee
where $\vek k_0$ is any constant vector, and $\vek j_0(\vek r)\neq 0$, \eqref{avwork} remains valid if the source contains a sufficiently narrow band of wavevectors around $\vek k_0$. This can be demonstrated by expressing the Fourier integrals of the source and the microscopic electric field, and averaging \eqref{workdens} over a unit cell. For more details on sources of finite sizes, see Appendix \ref{sec:finitesource}.

\section{Causality, passivity, and Kramers--Kronig relations}\label{sec:anal}
Here we will establish a framework for studying the analytic properties of the electromagnetic parameters, and the implications of passivity \cite{landau_lifshitz_edcm,agranovich84,dolgov81,alu11c,Yaghjian2013374}. If we use the Landau--Lifshitz formulation in which the medium is described solely by a permittivity $\vek\epsilon(\omega,\vek k)$, it has been stated that $\vek\epsilon(\omega,\vek k)$ is an analytic function in the upper half-plane $\im\omega>0$ for fixed $\vek k$, at least for sufficiently small $k$ \cite{landau_lifshitz_edcm,agranovich84}. This follows by regarding the electric field as the excitation and the displacement field as the response. However, as pointed out in Ref. \cite{dolgov81}, such an argument is not compelling since the electric field includes the response of the medium. Here we will use the relation between the applied source and the resulting field to prove that for fixed, real $\vek k$, the Landau--Lifshitz permittivity tensor $\vek\epsilon(\omega,\vek k)$ is analytic in the upper half-plane, even for anisotropic, bianisotropic, and spatially dispersive media. We will also provide the passivity condition.

Since the medium is assumed linear and time-shift invariant, the resulting macroscopic field $\vek E$ is related to the source $\vek J_\text{ext}$ by a linear relation
\be\label{ERJresp}
\vek E = \vek G(\omega,\vek k)\vek J_\text{ext},
\ee
where $\vek G(\omega,\vek k)$ is a (matrix) response function. For simplicity in notation we have suppressed the $\omega$ and $\vek k$ dependence of the fields. Recall that the medium is assumed to be causal and passive, so if the time-domain source current is any finite-duration pulse starting at $t=0$, the time-domain electric field vanishes for $t<0$ and does not blow up as $t\to\infty$. It follows that
\be\label{Ranalytic}
\vek G(\omega,\vek k) \text{ analytic for $\im\omega > 0$ and fixed $\vek k$.}
\ee
This applies to all elements of the matrix, since $\vek J_\text{ext}$ can be chosen to point in any direction.

Since the work done by the source must be nonnegative, we must have $-\re \vek J_\text{ext}^*\cdot\vek E\geq 0$ (see \eqref{avworkE}), or
\be\label{workineq}
-\re\vek J_\text{ext}^\dagger\vek G(\omega,\vek k)\vek J_\text{ext} \geq 0,
\ee
for real frequencies. Here $\dagger$ stands for hermitian conjugate, i.e., transpose and complex conjugate. We have argued for \eqref{Ranalytic} and \eqref{workineq} using a single $\vek k$ source; however as shown in Appendix \ref{sec:finitesource} they also follow when using a realistic source of finite size. Inequality \eqref{workineq} is valid in the upper half-plane $\im\omega>0$, as shown in Appendix \ref{sec:finitesource}.

Define a function 
\be
f(\omega)=-\vek J_0^\dagger\vek G(\omega,\vek k)\vek J_0,
\ee
where $\vek J_0$ is an arbitrary, but constant vector. We have just seen that $\re f(\omega)\geq 0$ for $\im\omega>0$. In fact, since $f(\omega)$ is an analytic function, it must be that $\re f(\omega)>0$ for $\im\omega>0$: Assume $f(\omega)=0$ somewhere in the upper half-plane. A zero of an analytic function is isolated, and in the vicinity of a zero, the function's complex argument takes all values from 0 to $2\pi$. This would make $\re f(\omega)<0$ somewhere around the zero, which contradicts $\re f(\omega)\geq 0$.

We have proved that
\be\label{Rpos}
-\re \left[\vek J_0^\dagger \vek G(\omega,\vek k)\vek J_0\right] > 0 \text{ for }\im\omega>0
\ee
for any, constant $\vek J_0$. Thus $\vek E=\vek G(\omega,\vek k)\vek J_\text{ext}\neq 0$ for any nonzero $\vek J_\text{ext}$, for $\im\omega>0$. This means that 
\be\label{Rinvertible}
\det\vek G(\omega,\vek k)\neq 0 \quad\text{for }\im\omega>0.
\ee
In other words, for all $\omega$ in the upper half-plane, we can invert $\vek G(\omega,\vek k)$ to obtain $\vek G(\omega,\vek k)^{-1}$. Since $\vek G(\omega,\vek k)$ is analytic, so is $\vek G(\omega,\vek k)^{-1}$. 

In the Landau--Lifshitz formulation, the Maxwell equations take the form
\begin{subequations}\label{maxwellLL}
\begin{align}
i\vek k\times\vek E-i\omega\vek B &=0, \\
\frac{1}{\mu_0} i\vek k\times\vek B + i\omega\epsilon_0\vek\epsilon(\omega,\vek k)\vek E &= \vek J_\text{ext}
\end{align}
\end{subequations}
in the frequency--wavenumber space. Combining them, we obtain
\be\label{maxwmatr}
\left(k^2\vek I_\perp-\frac{\omega^2}{c^2}\vek\epsilon(\omega,\vek k)\right)\vek E = i\omega\mu_0\vek J_\text{ext},
\ee
with $\vek I_\perp=\vek I - \vek k\vek k/k^2$ (or expressed by \eqref{Iperpx} in a coordinate system where $\vek k=k\vekh x$). Comparing \eqref{maxwmatr} and $\vek E=\vek G(\omega,\vek k)\vek J_\text{ext}$, we identify
\be\label{Rinv}
\vek G(\omega,\vek k)^ {-1} = (i\omega\mu_0)^{-1}\left(k^2\vek I_\perp-\frac{\omega^2}{c^2}\vek\epsilon(\omega,\vek k)\right).
\ee
We have already proved that $\vek G(\omega,\vek k)^{-1}$ is analytic in the upper half-plane $\im\omega>0$; thus so is $\vek\epsilon(\omega,\vek k)$.

With an asymptotic behavior $\epsilon(\omega,\vek k)\to\vek\epsilon(\infty,\vek k)$ as $\omega\to\infty$ and $\vek k$ is fixed, we can now state the Kramers--Kronig relations for $\vek\chi(\omega,\vek k)\equiv \vek\epsilon(\omega,\vek k)-\vek\epsilon(\infty,\vek k)$: 
\begin{subequations}\label{KK}
\begin{align}
& \re\vek\chi(\omega,\vek k) = \frac{2}{\pi}\text{P}\int_0^\infty \frac{\im\vek\chi(\nu,\vek k)\nu}{\nu^2-\omega^2}\diff\nu, \\
& \im\vek\chi(\omega,\vek k) = -\frac{2\omega}{\pi}\text{P}\int_0^\infty \frac{\re\vek\chi(\nu,\vek k)}{\nu^2-\omega^2}\diff\nu.
\end{align}
\end{subequations}
Here $\omega$ is real, and P stands for the Cauchy principal value. To obtain the Kramers--Kronig relations from the analyticity and the asymptotic behavior, we have used the Titchmarsh' theorem \cite{titchmarsh}. To this end we have assumed that $\vek\chi(\omega,\vek k)\to 0$ sufficiently fast as $|\omega|\to\infty$, and that $\chi(\omega,\vek k)$ does not have singularities for real frequencies \footnote{To establish Kramers--Kronig relations, the Titchmarsh' theorem requires the function to be uniformly square integrable along a line in the upper half-plane, parallel to the real axis. The assumption is for example valid if the function vanishes as $1/|\omega|$ or faster, but clearly, weaker conditions are possible. If the function has singularities on the real axis, modified Kramers--Kronig relations can be derived \cite{landau_lifshitz_edcm}}.

Substituting $\vek J_0=\vek G(\omega,\vek k)^{-1}\vek E_0$ in \eqref{Rpos} gives
\be\label{RposE}
-\re \left[\vek E_0^\dagger \vek G(\omega,\vek k)^{-1\dagger}\vek E_0\right] > 0 \text{ for }\im\omega>0,
\ee
valid for any vector $\vek E_0$. This can be written 
\be\label{passivitygenR}
-\vek G(\omega,\vek k)^{-1} - \vek G(\omega,\vek k)^{-1\dagger} \text{ positive definite},
\ee
or, using \eqref{Rinv},
\begin{align}
  i&\left[ \omega\vek\epsilon(\omega,\vek k)-\frac{k^2c^2}{\omega}\vek I_\perp \right]^\dagger 
 -i \left[ \omega\vek\epsilon(\omega,\vek k)-\frac{k^2c^2}{\omega}\vek I_\perp \right] \nonumber\\ 
 & \text{positive definite}. \label{passivitygen}
\end{align}
In principle, the passivity condition \eqref{passivitygen} has been derived for $\im\omega>0$. By taking the limit $\im\omega\to 0$, the passivity condition remains valid for all $\re\omega$ where this limit exist, provided we relax ``positive definite'' to ``positive semidefinite''. When both $\omega$ and $k$ are real, as is the case when they represent a Fourier component in time and space, the passivity condition becomes
\be\label{passivityreal}
-i\omega[\vek\epsilon(\omega,\vek k) - \vek\epsilon(\omega,\vek k)^\dagger] \text{ positive semidefinite}.
\ee
This reduces to the well known condition $\im\epsilon(\omega,\vek k) \geq 0$ for scalar permittivity and positive frequency.

\section{Source of finite size}\label{sec:finitesource}
In Appendix \ref{sec:anal} we imagined a source with a single wavevector $\vek k$. This source is somewhat unphysical, since it is present everywhere. Here we will consider sources of finite size, and re-derive the causality and work results \eqref{Ranalytic} and \eqref{workineq}. 

In this section we will write out the $\omega$ and $\vek k$ dependence explicitly. Let us consider a causal source in product form
\be
\vek J_\text{ext}(\omega,\vek k) = F(\vek k)\vek W(\omega).
\ee
From \eqref{floqebe} and \eqref{per} the frequency-domain microscopic electric field is
\begin{align}
\vek e(\omega,\vek r) &= \frac{1}{(2\pi)^3}\int \sum_{lmn} \vek E_{lmn}(\omega,\vek k)\e{i\vek b_{lmn}\cdot\vek r+i\vek k\cdot\vek r}\diff^3k \nonumber\\
&= \frac{1}{(2\pi)^3}\sum_{lmn} \int \vek E_{lmn}(\omega,\vek k)\e{i\vek b_{lmn}\cdot\vek r+i\vek k\cdot\vek r}\diff^3k \nonumber\\
&= \frac{1}{(2\pi)^3}\sum_{lmn} \int \vek E_{lmn}(\omega,\vek k'-\vek b_{lmn})\e{i\vek k'\cdot\vek r}\diff^3k' \nonumber\\
&= \frac{1}{(2\pi)^3}\int \sum_{lmn} \vek E_{lmn}(\omega,\vek k-\vek b_{lmn})\e{i\vek k\cdot\vek r}\diff^3k,
\end{align}
where $\vek E_{lmn}(\omega,\vek k)$ is proportional to $F(\vek k)$. Since $\vek e(\omega,\vek r)$ describes a causal field for all $\vek r$, we must have
\be\label{sumsource}
\sum_{lmn} \vek E_{lmn}(\omega,\vek k-\vek b_{lmn}) \text{ causal, for fixed $\vek k$.}
\ee
The source function $F(\vek k)$ can be chosen such that 
\begin{align}
& F(\vek k_0)=1 \text{ for } lmn = 000, \nonumber\\
& F(\vek k_0-\vek b_{lmn})=0 \text{ for all } lmn\neq 000.
\end{align}
In other words there is a peak at $\vek k=\vek k_0$, and zeros at $\vek k=\vek k_0-\vek b_{lmn}$ for $lmn\neq 000$. This is achieved e.g. if 
\be\label{Asource}
F(\vek k)=\text{sinc}^2[(k_x-k_{0x})a]\text{sinc}^2[(k_y-k_{0y})a]\text{sinc}^2[(k_z-k_{0z})a].
\ee
This source has a finite extent, as seen by inverse Fourier transforming \eqref{Asource}. By setting $\vek k=\vek k_0$, and considering \eqref{sumsource}, we find that $\vek E(\omega,\vek k)\equiv \vek E_{000}(\omega,\vek k)$ is causal. 

We can write 
\be\label{respEGJ}
\vek E(\omega,\vek k) = \vek G(\omega,\vek k)\vek J_\text{ext}(\omega,\vek k),
\ee
where $\vek G(\omega,\vek k)$ is a (tensor) response function. We choose a source with finite duration in the time-domain. Since the medium is passive, the electric field does not blow up as $t\to\infty$. Since $\vek J_\text{ext}(\omega,\vek k)$ and $\vek E(\omega,\vek k)$ are causal, it follows that they are analytic in the upper half-plane $\im\omega>0$. As $\vek J_\text{ext}(\omega,\vek k)$ is otherwise arbitrary, the response function $\vek G(\omega,\vek k)$ must therefore be analytic for $\im\omega>0$ for each fixed $\vek k$ \eqref{Ranalytic}.

The properties of $\vek G(\omega,\vek k)$ in the upper half-plane can be further explored by considering sources with time dependence $\exp(\gamma t-i\omega't)$ \cite{Brune31}:
\be
\vek j_\text{ext}(t,\vek r) = \re \left[\vek f(\vek r) u(t) \e{\gamma t-i\omega't} \right].
\ee
Here $u(t)$ is the unit step function, $\gamma>0$, and $\omega'$ is real. Taking $\vek f(\vek r)$ to be real, this source can be expressed in frequency--wavenumber space
\be
\vek J_\text{ext}(\omega,\vek k) = \vek F(\vek k)W(\omega),
\ee
where $\vek F(\vek k)$ is the Fourier transform of $\vek f(\vek r)$, and $W(\omega)$ is the Laplace transform of $\e{\gamma t}\cos(\omega' t)$, after setting the Laplace variable $s=-i\omega$. 

At least for $t\gg 1/\gamma$, the transients can be ignored compared to the exponentially increasing field. Then the electric field will be of the form $\exp(\gamma t-i\omega't)$, and the power density $p_\text{ext}(t) = -\vek j_\text{ext}(t,\vek r)\cdot\vek e(t,\vek r)$ becomes
\begin{align}
& p_\text{ext}(t) = -\re\left( \frac{\e{\gamma t-i\omega't}}{(2\pi)^3}\int \vek F(\vek k)\e{i\vek k\cdot\vek r}\diff^3 k \right)  \label{powgamma}\\
& \cdot \re\left( \frac{\e{\gamma t-i\omega't}}{(2\pi)^3} \int\sum_{lmn}\tilde{\vek E}_{lmn}(\omega,\vek k-\vek b_{lmn})\e{i\vek k\cdot\vek r}\diff^3k \right). \nonumber
\end{align}
Here $\tilde{\vek E}_{lmn}(\omega,\vek k)$ is the same as $\vek E_{lmn}(\omega,\vek k)$ except that the factor $W(\omega)$ has been removed, and $\omega=\omega'+i\gamma$. Using $\re\alpha=(\alpha+\alpha^*)/2$, and integrating the resulting expression over all space and from time $t_0$ to $t_1$, we find the total work in this time interval:
\begin{align}
& W_\text{ext} = \int_{t_0}^{t_1}\int p_\text{ext}\diff^3 r \diff t\\
&= -\frac{\e{2\gamma t_1}-\e{2\gamma t_0}}{4\gamma(2\pi)^3} \re\int\vek F^*(\vek k)\cdot\sum_{lmn}\tilde{\vek E}_{lmn}(\omega,\vek k-\vek b_{lmn})\diff^3 k \nonumber\\
&+ \re \left[C\cdot(\e{2\gamma t_1-2i\omega' t_1}-\e{2\gamma t_0-2i\omega' t_0})\right], \nonumber
\end{align}
where $C$ is a complex-valued quantity which is independent of $t_0$ and $t_1$. Let $t_0\gg 1/\gamma$. Since $t_0$ is finite, the source has only done a finite amount of work $W_0$ before $t_0$. Assuming the medium has no stored energy before $t=0$, we must have $W_0+W_\text{ext}\geq 0$. For a sufficiently large $t_1$, but such that $C\e{-2i\omega't_1}$ is imaginary, we obtain the condition
\be \label{workgammat1}
-\re\int\vek F^*(\vek k)\cdot\sum_{lmn}\tilde{\vek E}_{lmn}(\omega,\vek k-\vek b_{lmn})\diff^3 k \geq 0.
\ee
Recall that $\tilde{\vek E}_{lmn}(\omega,\vek k)$ is proportional to $F(\vek k)$. Thus, by picking a source with a sufficiently narrow, effective band $\Delta k$ around a fixed wavenumber $\vek k_0$ ($\Delta ka\ll 1$, which means that the source must cover several unit cells), we can make the terms with $lmn\neq 000$ arbitrarily small. Hence we must have 
\be \label{workgammat2}
-\re\int \vek F^*(\vek k)\cdot\tilde{\vek E}(\omega,\vek k)\diff^3 k \geq 0.
\ee
We now use \eqref{respEGJ}, which means $\tilde{\vek E}(\omega,\vek k) = \vek G(\omega,\vek k)\vek F(\vek k)$. Picking a $\vek F(\vek k)$ which is narrow banded in $\vek k$ compared to the variations in $\vek G(\omega,\vek k)$, we obtain
\be \label{workR}
-\re \vek J_0^\dagger \vek G(\omega,\vek k)\vek J_0 \geq 0
\ee
for all constant vectors $\vek J_0$.

\section{Analyticity of tensor elements}\label{sec:analtensor}
Suppose we have an expansion in the form
\be\label{finitetayl}
f(\omega,\vek k) = a(\omega) + b_i(\omega)k_i + c_{ij}(\omega)k_ik_j,
\ee
where $a(\omega)$, $b_i(\omega)$, and $c_{ij}(\omega)$ are independent of $\vek k$. We take $c_{ij}(\omega)$ to be symmetric, as any antisymmetric part is irrelevant for the expansion. Let $f(\omega,\vek k)$ be an analytic function of $\omega$ (in a given domain), for any fixed $\vek k$. We will prove that the coefficients $a(\omega)$, $b_i(\omega)$, and $c_{ij}(\omega)$ are analytic.

By putting $\vek k=0$, we find that $a(\omega)=f(\omega,0)$ is analytic. Considering 
\be
f(\omega,\vek k)-f(\omega,-\vek k) = 2b_i(\omega)k_i,
\ee
it follows that $b_i(\omega)$ is analytic. We now have that 
\be
c_{ij}(\omega)k_ik_j = f(\omega,\vek k) - a(\omega) - b_i(\omega)k_i
\ee
is analytic. By letting $\vek k$ point in the $\vekh x$, $\vekh y$ or $\vekh z$ direction, we find that $c_{ii}(\omega)$ are analytic for any $i$. Finally we obtain e.g. that $c_{12}$ is analytic by letting $\vek k=k(\vekh x+\vekh y)/\sqrt 2$.

The argument can be extended to an infinite Taylor series, or a series with a remainder term, by noting that the partial derivatives $\partial f/\partial k_i$ and $\partial^2 f/\partial k_i\partial k_j$ are analytic. This follows by using the Cauchy--Riemann equations, assuming symmetry of second order derivatives.

In particular, if the Landau--Lifshitz permittivity is expressed in the form
\be\label{epsLL1211}
\epsilon_{ij}(\omega,\vek k) - \delta_{ij} = \chi_{ij} + \alpha_{ikj}k_k/\epsilon_0 + \beta_{iklj}k_kk_l c^2/\omega^2, 
\ee
the analyticity of $\epsilon_{ij}(\omega,\vek k)$ means that the tensors $\chi_{ij}$, $\alpha_{ikj}$, and $\beta_{iklj}$ are analytic.

\def\cprime{$'$} \def\cprime{$'$}
%

%\bibliography{nbib,nbibHOH}% Produces the bibliography via BibTeX.
\end{document}